\documentclass[12pt,preprint]{emulateapj}

\shorttitle{Echelle diagram and rotation for the sample}
\shortauthors{Papar\'o et al.}

\begin{document}

\title{Unexpected series of regular frequency spacing of $\delta$ 
Scuti stars in the non-asymptotic regime -- II. Sample -- echelle diagrams and rotation
}

\author{M. Papar\'o\altaffilmark{1}, J.~M. Benk\H{o}\altaffilmark{1}, 
M. Hareter\altaffilmark{1}, and J.~A. Guzik\altaffilmark{2}} 

\altaffiltext{1}{Konkoly Observatory, MTA CSFK, Konkoly Thege Mikl\'os \'ut 15-17., H-1121 Budapest, Hungary}
\altaffiltext{2}{Los Alamos National Laboratory, Los Alamos NM 87545 USA}

\email{paparo@konkoly.hu}

\begin{abstract}

A sequence search method was developed for searching for regular frequency spacing 
in $\delta$ Scuti stars by visual inspection and algorithmic search. The sample contains 
90 $\delta$ Scuti stars observed by CoRoT. An example is given to represent 
the visual inspection. The algorithm (SSA) is described in detail. The data treatment of the CoRoT 
light curves, the criteria for frequency filtering and the spacings derived by two methods 
(three approaches: VI, SSA and FT) are given for each target. 
Echelle diagrams are presented for 77 targets, for which 
at least one sequence of regular spacing was identified.
Comparing the spacing and the shifts between pairs of echelle ridges revealed that 
at least one pair of echelle ridges is shifted to midway between the spacing for 22 stars. 
The estimated rotational frequencies compared to the shifts revealed rotationally split doublets, 
triplets and multiplets not only for single frequencies, but for the complete echelle 
ridges in 31 $\delta$ Scuti stars. 
Using several possible assumptions for the origin of the spacings, we derived the 
large separation ($\Delta\nu$), which are distributed along the mean density versus large separations 
relation derived from stellar models \citep{Suarez14}.
\end{abstract}


\section{Introduction}

Delta Scuti stars are very complex pulsators. They are located on and above the main sequence,
 they pulsate mainly in p-type and g-type non-radial modes, beside the radial ones. The modes 
are excited by the $\kappa$-mechanism in the He ionization zone (\citealt{Unno81, Aerts10}). 
The amplitudes of the radial modes are remarkably lower than in the classical radial pulsators, 
although they lie in the extension of the classical instability strip to the main sequence. 
They are close to the Sun on the HR diagram, but due to the excitation of the low order modes, 
no high level regularity of the modes has been predicted among them. 
Classical pulsators, with simple structure of the excited modes, and the Sun, with stochastically 
excited high-order modes that are predicted to have regular frequency spacing, 
have the advantage for mode identification.

The space missions yielded the detection of a huge number of $\delta$ Scuti stars with a much higher 
signal to noise ratio than we had before (\citealt{Baglin06, Auvergne09, Borucki10}). 
They allowed us the detection of a much larger set of modes. 
In the era of ground-based observations, we had hoped to match the increased number of modes 
by comparing them directly to model frequencies. Unfortunately this hope has not been realized due 
to the still existing discrepancy between the numbers of observed and predicted frequencies.

Up to now we could not avoid the traditionally used methods of mode identification, 
using the color amplitude ratio and phase differences 
\citep{Watson88, Viskum98, Balona99, Garrido00}.

The basic problem in mode identification of $\delta$ Scuti stars 
is the rotational splitting of modes due to intermediate and fast rotation. 
Starting from the first-order effect in slow rotators \citep{Ledoux51}, the second-order 
effects \citep{Vorontsov81, Vorontsov83, Dziembowski92} and the third-order effects 
\citep{Soufi98} were intensively investigated theoretically in the frame of the perturbative 
theory and were applied for individual stars \citep{Templeton00, Templeton01, Pamyatnykh98}.

The theoretical investigation of the intermediate and fast rotating stars exhibited 
a rapid improvement since the work of \citet{Lignieres06} and \citet{Roxburgh06}. In the following years,
a series of papers \citep{Lignieres08, Lignieres09,Lignieres10, Reese08, Reese09} investigated
different aspects of the ray dynamic approach for fast rotating stars. Instead of the traditional 
quantum numbers ($l$, $n$), they introduced the modified quantum numbers ($\hat{l}$, $\hat{n}$) 
including the odd and even parity of modes in fast rotating stars. They reached a level that 
recently echelle diagrams were published; for example, see \citet{Ouazzani15}.

In the ray dynamic approach different families of modes, named the low frequency modes, 
whispering gallery modes, chaotic modes and island modes were recognized. These modes represent 
different pulsational behavior. The low frequency modes are counterparts of the high-order g-modes. 
They have negligible amplitude in the outer layers, so they should not be detected observationally. 
The whispering gallery modes are counterparts of the high degree acoustic modes. 
They probe the outer layers but due to low visibility they might not be detected. Chaotic modes 
do not have counterparts in the non-rotating case. 
Due to the lack of symmetry in the cancellation and the significant amplitude in the whole of the 
stellar interior, these modes are expected to be detected observationally. However, 
they appear only in very fast rotating models. Island modes are counterparts of the low degree 
acoustic modes. They probe the outer layers of the star and present good geometric visibility. 
Therefore these modes should be easily detected observationally. Low $\hat{l}$ modes are expected to 
be the most visible modes in the seismic spectra of rapidly rotating stars.
For a given parity, the mode frequencies line up along ridges of given $\hat{l}$ values. 
However, the first difficulty with studying the island modes is to be able to identify them 
among all the other type of modes present in the spectrum of rapidly rotating stars 
(chaotic and whispering gallery modes).

The regular arrangement of the excited modes in stars having high order p modes 
(Sun and solar-type oscillation in red giants) or having high order g modes (white dwarfs) allowed 
us to reach the asteroseismology level. The radial distribution of the physical parameters 
(pressure, temperature, density, sound speed and chemical composition) were derived for the Sun. 
The mode trapping allowed us to derive the masses of the H and He layers in white dwarfs.

Using the space data many investigations aimed to find regularity in the $\delta$ Scuti 
stars, in MOST data \citep{Matthews07}, in CoRoT data \citep{Garcia Hernandez09, Garcia Hernandez13, Mantegazza12} and in {\it Kepler} data
\citep{Breger11, Kurtz14}. The most comprehensive study \citep{Garcia15} reported 
regularities for 11 stars on a sample of 15 {\it Kepler} $\delta$ Scuti stars, providing the 
large separation for them. They revealed two echelle ridges with 6 and 4 frequency members 
for KIC 1571717. Up to now this is the most extended survey for regularities in $\delta$ Scuti stars.

Our goal was to survey the possible regularities of $\delta$ Scuti stars on a much 
larger sample of CoRoT targets. In addition, as a new method we searched for complete 
sequence(s) of quasi-equally spaced frequencies with two approaches, namely visual inspection 
and algorithmic search. We present in this paper our detailed results for the whole sample.

\section{CoRoT data}\label{data}

The CoRoT satellite was launched in 2006 \citep{Baglin06}. LRa01, the first long run in 
the direction of anti-center, started on October 15, 2007 and finished on
March 03, 2008, resulting in a $\Delta$T=131 day time span. Both chromatic and
monochromatic data were obtained on the EXO field with a regular sampling of 8 minutes, 
although for some stars an oversampling mode (32s) was applied. After using the CoRoT pipeline 
\citep{Auvergne09} the reduced N2 data were stored in the CoRoT data archive. 
Any kind of light curves of the EXO field were systematically
searched for $\delta$ Scuti and $\gamma$ Doradus light curves by one of us \citep{Hareter13}. 
 
We did not rely on the automatic classification tool (CVC, \citealt{Debosscher09}) because of ambiguities 
and the risk of misclassifications that might have appeared in the original version. 
Rather, we selected the targets by visual inspection of light curves and their Fourier transform 
and kept those for which classification spectra (AAOmega, \citealt{Guenther12, Sebastian12}) were available.
A recent check of the new version of CVC
(CoRoT N2 Public Archive\footnote{\url{http://idoc-corotn2-public.ias.u-psud.fr/invoquerSva.do?sva=browseGraph}}, 
updated 2013 February) revealed that most of our stars (57) were classified 
as $\delta$ Scuti stars with high probability. 
Some GDOR (4), MISC (11), ACT (5) and $\beta$ Ceph (3) classifications also appeared.

The initial sample of our investigation consists of 90 $\delta$ Scuti stars 
extracted from the early version of N2 data in the archive. Nowadays a modified 
version of N2 data on LRa01 can be found in the archive. Comparing our list and the new version, 
we noticed that the light curve of 14 stars from our initial 
sample had been omitted from the new version. The low peak-to-peak amplitude of the light curve, 
in some cases, might explain the decision but we did not find any reasons why targets with 
peak-to-peak amplitude from 0.01 to 0.05 mag had been excluded. We therefore kept these stars in our initial sample.

Because the CoRoT N2 data are still affected by several instrumental effects,
we used a custom IDL-code that removes the outliers and corrects for jumps and trends. 
The jumps were detected by using a two sampled t-test with sliding 
samples of 50 data points and the trends were corrected by fitting low order 
polynomials. The outliers were clipped using an iterative median filter, 
where a 3$\sigma$ rejection criterion was employed. The range of the light variation for most of 
the stars is 0.003 - 0.04 magnitude, with the highest population around 0.01 mag. 
The brightness range is from 12.39 to 15.12 mag, covering almost three magnitudes.

The frequencies were extracted using the software SigSpec {\citep{Reegen07} 
in the frequency range from 0 to 80 d$^{-1}$. The significance limit was set initially
to 5. The resulting list of frequencies for 90 $\delta$ Scuti stars served as an 
initial database for our frequency search \citep{Hareter13}.

\begin{deluxetable}{rrrr}
\tablecaption{List of excluded targets \label{omit}}  
\tablehead{
\colhead{No} & \colhead{CoRoT ID} & \colhead{SSF} & \colhead{FF} }
\startdata
16 & 102713193 & 52 & $-$ \\
17 & 102614844 & 78 & $-$ \\
42 & 102646094 & 45 & $-$ \\
44 & 102746628 & 51 & $-$ \\
57 & 102763839 & 93 & $-$ \\
58 & 102664100 & 35 & $-$ \\
59 & 102766985 & 61 & $-$ \\
60 & 102668347 & 123 & $-$ \\
61 & 102668428 & 57 & $-$ \\
64 & 102706982 & 68 & $-$ \\
\tableline
41 & 102645677 & 106 & 14 \\
46 & 102749985 & 63 & 9 \\
85 & 102589213 & 70 & 10 
\enddata                                               
\tablecomments{The columns contain the running numbers (No), official CoRoT ID, the number of SigSpec 
frequencies (SSF), and the number of filtered frequencies (FF), respectively.}
\end{deluxetable} 

\subsection{The final sample of targets and filtering}

We filtered the SigSpec frequencies using some trivial ideas (tested for CoRoT data 
by \citealt{Balona14}). We removed

\begin{itemize}

\item
low frequencies close to 0 d$^{-1}$ in most cases up to 2 d$^{-1}$, since we were primarily 
interested in the $\delta$ Scuti frequency region

\item
the possible technical peaks connected to the orbital period of the spacecraft
($f_{\mathrm {orb}}$= 13.97 d$^{-1}$)

\item 
frequencies of lower significance in groups of closely spaced peaks, because they 
are most likely due to numerical inaccuracies during the pre-whitening cascade. 
We kept only the highest amplitude ones.

\item

the low-amplitude, low-significance frequencies in general.
The lowest amplitude limit was different from star to star, since the frequencies 
showed a different amplitude range from star to star, but it was around 0.1 mmag in general.
\end{itemize}

 We might dismiss true pulsating modes in the filtering process, 
but finding regularities among fewer frequencies is more convincing. Accidental 
coincidences could appear with higher probability if we use a larger set of frequencies. 
After finding a narrow path in solving the pulsation-rotation connection we may widen the 
path to a road.

In 10 stars only a few frequencies remained after the filtering process. In each case a 
dominant peak remained in the $\delta$ Scuti frequency range giving an excuse of the 
positive classification as a $\delta$ Scuti star.
The limited number of frequencies in these stars was not enough for our main purpose 
(to find regularities between the frequencies), so we omitted them from further investigation. 

In addition, we did not find any regularities in three stars. The 13 stars, that were omitted for 
any reason, are listed in Table~\ref{omit}.
Our finally accepted sample where we found regularities with one of our methods is listed in Table~\ref{bigtable}.
In both tables the CoRoT ID of the stars is given in the second column. 
For the sake of simpler treatment during the investigation we introduced a running number  
(first column in the tables). We refer to the stars by the running number in the rest of the paper. 
The 96 running numbers instead of 90 are due to a special test checking the ambiguity of our results. 
The double running numbers mean stars (see CoRoT ID in Table~\ref{bigtable}) where the filtering of 
SigSpec frequencies and the search for periodic spacing were independently done for the same stars (6 stars). The running numbers representing the same stars were identified (connected to each other) only at the end of the searching process. The independent cleaning due to the not-fixed limiting amplitude and subjectivity resulted in different numbers of the frequencies and consequently in different values of the spacing, the number of the frequencies in the echelle ridges, and the numbers of 
echelle ridges.
The number of independent $\delta$ Scuti stars in our sample is 77, where we got positive results with 
one of our methods concerning the regular spacing. The $T_{\mathrm{eff}}$, $\log g$ and radial 
velocity ($v_{\mathrm {rad}}$), derived by one of us \citep{Hareter13} are presented in the third, fourth and fifth column of 
Table~\ref{bigtable}.

\begin{figure*}
\includegraphics[width=17cm]{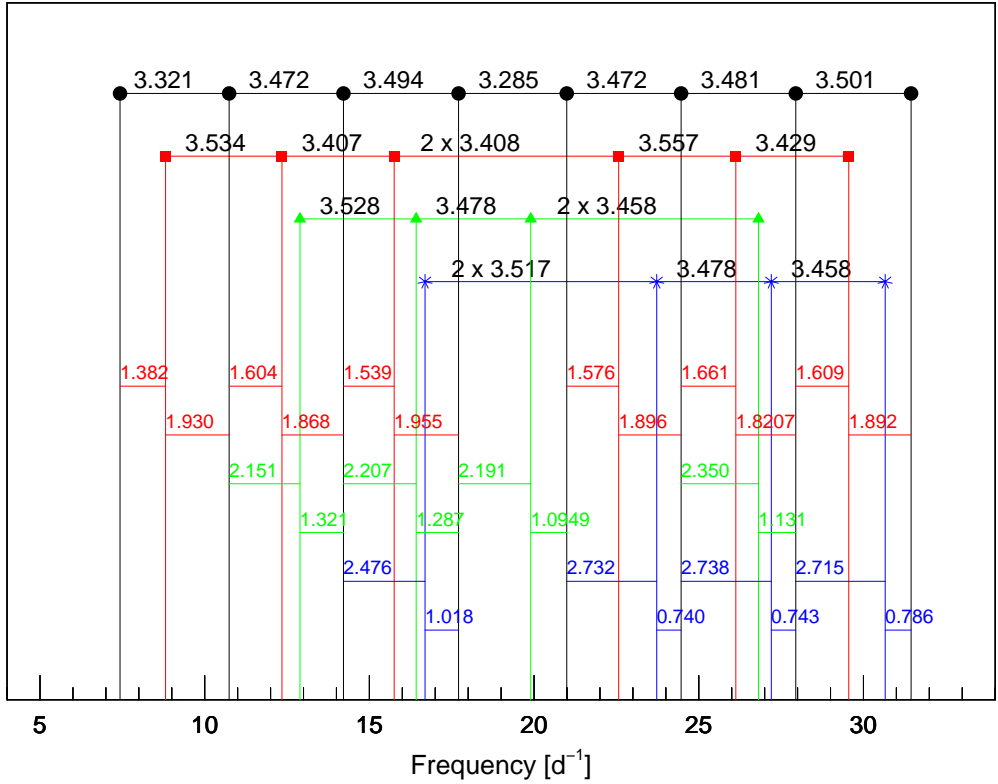}
\caption[]{
Sequences with quasi-equal spacing, and shifts of the sequences for star 
No. 65 (CoRoT 102670461). 1st -- black dots, average spacing 3.431$\pm$0.091 d$^{-1}$; 
2nd -- red squares, 3.467$\pm$0.073 d$^{-1}$; 3rd -- green triangles, 3.488$\pm$0.036 d$^{-1}$; 
4th -- blue stars, 3.484$\pm$0.030 d$^{-1}$ average spacings were obtained. 
The mean spacing of the star is 3.459$\pm$0.030 d$^{-1}$. 
The shifts of the 2nd, 3rd and 4th sequences relative to the first one are also given in the same color as the sequences.
} \label{fig1}
\end{figure*}

The filtering  guidelines yielded a much reduced number of frequencies. 
For comparison we listed the number of SigSpec frequencies (SSF) and that of 
the filtered frequencies (FF) in the 6th and 7th columns.
Only about 20-30\% of the original peaks were kept in our final list of frequencies. 
When the effectiveness of our method for finding regularities has been confirmed, 
the application could be extended including frequencies at lower amplitude.
For possible additional investigation we attached the filtered frequencies of each 
star to this paper in an electronic version\footnote{See the web page of this journal: \url{http://}}. 
Table~\ref{sample_data} shows an excerpt from this data file as an example. 
Additional information on flags is discussed later. 

\section{Search for periodic spacing}\label{data_processing}

\begin{deluxetable*}{ccrcccccc}
\centering
\tablecaption{Sample from the data file \label{sample_data}}
\tablehead{
\colhead{No} & \colhead{CoRoT ID}  &   \colhead{$f$} & \colhead{$A(f)$} &   
\colhead{VI1} &  \colhead{VI2} &  \colhead{SSA1} & \colhead{SSA2} & \colhead{SSA3} \\
 \colhead{}  &  \colhead{} &   \colhead{(d$^{-1}$)}  & \colhead{(mmag)} & 
\colhead{} &  \colhead{}    &  \colhead{} &  \colhead{} &  \colhead{}    }
\startdata
1  &  102661211 &    10.0232 & 8.462 &  0  &  $-$ &   1 &   1 &  $-$  \\
1  &  102661211 &     7.8170 & 3.606 &  2  &  $-$ &   2 &   5 &  $-$  \\
1  &  102661211 &    14.7389 & 1.990 &  3  &  $-$ &   6 &   0 &  $-$  \\
1  &  102661211 &    12.0054 & 1.602 &  6  &  $-$ &   2 &   4 &  $-$  \\
1  &  102661211 &     8.7854 & 1.437 &  5  &  $-$ &   4 &   0 &  $-$  \\
$\cdots$   &    $\cdots$   &   $\cdots$   &   $\cdots$   &   $\cdots$ 
&    $\cdots$   &    $\cdots$ & $\cdots$   & $\cdots$   
\enddata
\tablecomments{
This table is published in its entirety in the electronic edition of ApJS,
a portion is shown here for guidance regarding its form and content.
The columns contain local id, CoRoT ID, used frequency, Fourier amplitude of the frequency, and
 echelle ridge flags of the frequency obtained from the different search methods (VI or SSA), respectively.
The 0 value means that the frequency is not on any echelle ridges, while sign $-$ denotes nonexistent 
search result. See the text for details.}
\end{deluxetable*}

Investigations on the regular behavior of frequencies in $\delta$ Scuti stars and derivation of the 
large separation have been carried out in the past (see in \citealt{Paparo15}). 
Even in the earlier years clustering of non-radial modes around the frequencies of radial modes 
over many radial orders has been reported for a number of $\delta$ Scuti stars: 44 Tau, BL Cam, FG Vir 
\citep{Breger09}, giving the large separation. The 
clustering supposes that the sequence of low-order $l$=1 modes, slightly shifted with respect 
to the frequency of the radial modes, also reveals the large separation in the mean value \citep{Breger99}.
In all cases the histogram of the frequency differences or the Fourier Transform (FT) using the 
frequencies as input data were used. Both methods are sensitive to the most probable spacing frequency 
differences.

We searched for sequence(s) among the frequencies with quasi-equal spacing in our sequence search method. 
The visual inspection (VI) of our targets in the whole sample led us to establish the constraints for 
the Sequence Search Algorithm (SSA).
We present here the description of both the VI and the SSA and the results for the individual targets.

\subsection{Visual inspection (VI)}\label{visual}

In the visual inspection of the frequency distribution of our target, we recognized that 
almost equal spacing exists between the pair(s) of frequencies of 
the highest amplitude. The pairs proved to be connected to each other producing a sequence. 
New members with frequencies of lower amplitude were intentionally searched, so the 
sequence was extended to both the lower and higher frequency regions. Following the process with other 
pairs of frequencies of higher amplitude, we could localize more than one sequence, sometimes many sequences in a star. 
We noticed such an arrangement from star to star over the whole sample. We present here another 
example of the sequences compared to \cite{Paparo15} (paper Part I), to show how equal 
the spacings are between the members of a sequence, how the sequences are arranged compared to each other, and how we find a sequence if one consecutive member is missing. 
A new parameter appears in this process, namely the shifts of a frequency (member) to the consecutive 
lower and higher frequencies of the reference sequence (the first one is accepted). 

Fig.~\ref{fig1} shows four sequences of similar regular spacing for CoRoT 102670461 
(running number: 65). The sequences consist of 8, 6, 4 and 4 members, 
respectively, altogether including more than 45\% of the filtered frequencies. 
We allowed to miss one member of the 
sequence, if the half of the second consecutive member's spacing matched the 
regular spacing. In this particular case the missing members of the sequences are in the 20-23.5 d$^{-1}$ interval, which is in general the middle of the interval of the usually excited modes in $\delta$ Scuti stars. The frequencies of the highest amplitudes normally appear in this region. The mean value of the spacing is 
independently given for each sequence in the figure's caption.
The mean values differ only in the second digits. The general spacing value, 
calculated from the average of the sequences, is 3.459$\pm$0.030 d$^{-1}$. 

Fig.~\ref{fig1} also displays the shifts that we discussed before. 
They do not have random value, but represent characteristic values. 
Although the shifts are not the same for each member in a sequence, their mean 
values are characteristic for each sequence. We got 1.562$\pm$0.097 and 1.894$\pm$0.047 for the second, 
2.225$\pm$0.087 and 1.208$\pm$0.112 for the third and 2.665$\pm$0.127 
and 0.821$\pm$0.132 d$^{-1}$ for the fourth sequence relative to the first (reference) sequence.
The frequencies of the second sequence are almost 
midway between the first sequence, which we would expect in a comb-like structure. 
The third sequence is shifted by 0.635 d$^{-1}$ relative to the second one, while the fourth one is shifted by 0.297 d$^{-1}$ relative to the third one (practically half of the shift between the second and third sequences) although this value is determined only by averaging two independent values due to the missing members in the sequences. The shift of the fourth sequence relative to the second one is 1.069 d$^{-1}$.

 According to the AAO spectral classification \citep{Guenther12, Sebastian12},
CoRoT 102670461 has
 $T_{\mathrm {eff}}$=7325$\pm$150 K, $\log g$=3.575$\pm$0.793 and A8V spectral type and a variable
star classification as a $\delta$ Scuti type star \citep{Debosscher09}.
Following the process used by \citet{Balona15} for {\it Kepler} stars (discussed later in detail), we derived a possible
equatorial rotational velocity (100 km~s$^{-1}$) and a first-order rotational splitting (0.493 d$^{-1}$). Knowing the rotational splitting another regularity appears. 
The shift of the fourth sequence relative to the second one (1.069 d$^{-1}$) remarkably agrees with twice the value of the estimated equatorial rotational splitting. 
The appearance of twice the value of the rotational frequency is predicted by the theory \citep{Lignieres10}.

\begin{figure}
\includegraphics[width=9cm]{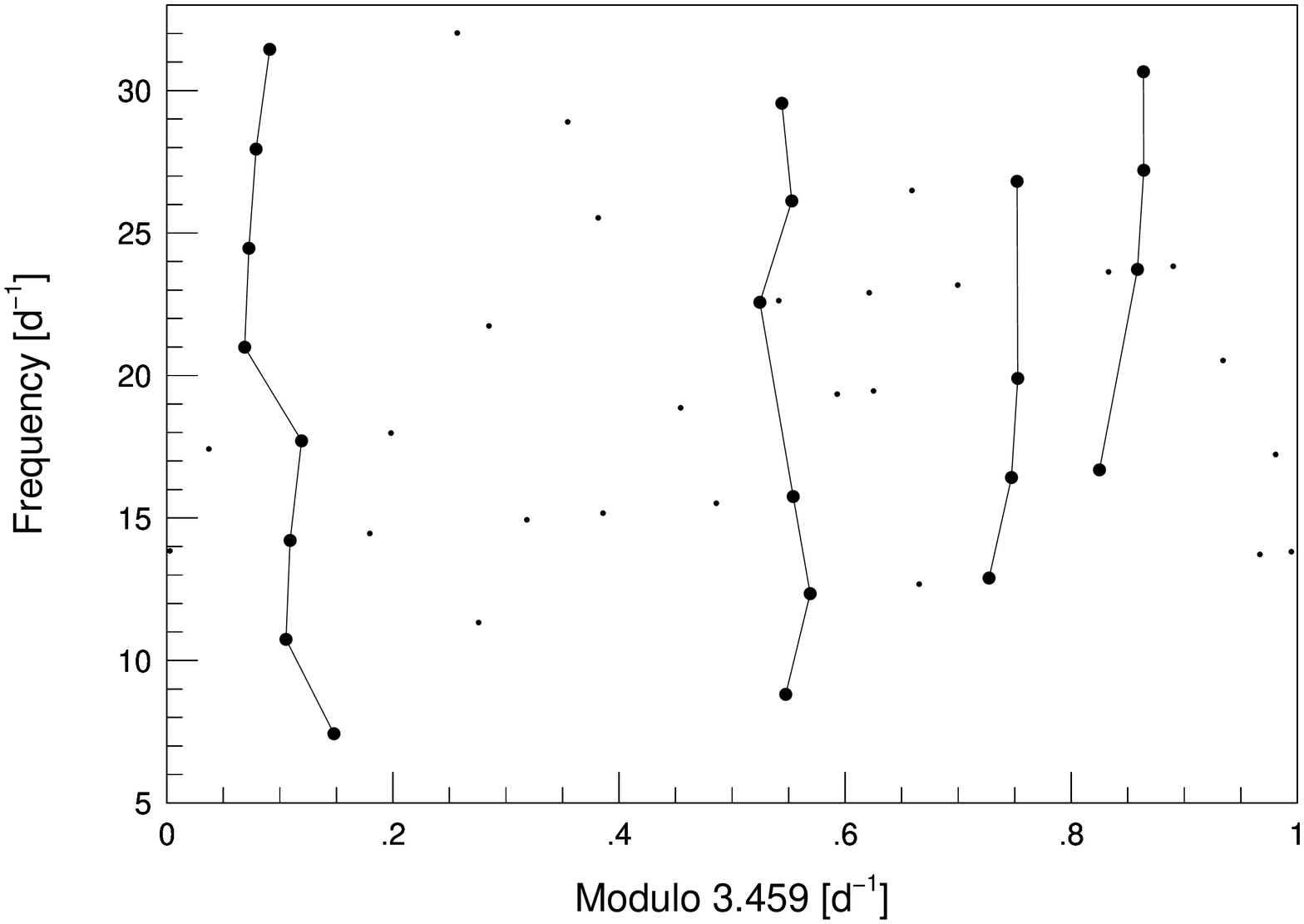}
\caption[]{
Echelle diagram of the star No. 65, consistent with the 
four sequences of Fig.~\ref{fig1}. 45\% of the filtered frequencies are located on the echelle ridges
(the other frequencies are shown by small dots).
} \label{fig2}
\end{figure}

The sequences in Fig.~\ref{fig1} are practically a horizontal representation of the widely used echelle diagram. 
In Fig.~\ref{fig2} we present the echelle diagram of the star No. 65, modulo 3.459 d$^{-1}$, 
in agreement with Fig.~\ref{fig1}. Fig.~\ref{fig2} displays all the filtered frequencies (small and large dots) but 
only 45\% of them are located on the echelle ridges (large dots).
The first, second, third and fourth sequences in the order of Fig.~\ref{fig1} agree with the echelle ridges at about 0.1, 0.55, 0.75 and 0.85 d$^{-1}$, respectively, in modulo value.

We found sequence(s) in 65 independent targets by the visual inspection. The number of
frequencies included in the sequences (EF$_{\mathrm {VI}}$), the number of sequences 
(SN$_{\mathrm {VI}}$) and the spacings (SP$_{\mathrm {VI}}$) 
are given in the 8th, 9th and 10th columns of Table~\ref{bigtable}, respectively. 
To archive the work behind these columns we added flags in five additional columns to 
the frequencies of the sequences in the electronic table (see also Table~\ref{sample_data}). 
Concerning a given star, many columns contain flags, as many spacing values were found by our methods. VI means the sequence of the visual inspection, while $1,2,\dots$, flag means that 
this frequency belongs to the 1st, 2nd, $\dots$, sequence. 
The $0$ flag marks those frequencies that are not located in any sequence. If the visual inspection 
resulted in more than one spacing, then VI1 and VI2 columns were filled in.
Using the flagged frequencies a similar diagram could be prepared for all targets, 
obtaining the individual spacings and the shifts which we presented in Fig.~\ref{fig1} for the star No. 65.
The distribution of the spacing obtained by visual inspection on the whole sample 
shows two dominant peaks between 2.3-2.4 d$^{-1}$ (10 stars) and an equal population between 3.2-3.5 d$^{-1}$ (7, 7 and 6 stars in each 0.1 d$^{-1}$ bins).

We summarize the results on the independent spacings as follows. 
The ambiguity of the personal decision is shown by six cases (double
numbering), where both the filtering process and the visual inspection were independently done. 
The most serious effect probably was the actual personal condition of the investigator. 
Since we do not want to polish our method we honestly present the differences in the solution.
Different EF$_{\mathrm {VI}}$, SN$_{\mathrm {VI}}$ and sometimes different spacing (SP$_{\mathrm {VI}}$) 
values were derived. However, in half of 
the cases the independent investigation resulted in similar spacing (stars No. 1=55, 2=66 and 8=92). 
In two stars one of the searches had negative results (stars No. 81 and 13) while the other search 
was positive (stars No. 11 and 74). There was only one case (star No. 14=96) where a completely 
different spacing value was obtained (1.844 versus 2.429, 3.3387 d$^{-1}$).
In a few cases (stars No. 50, 54 and 77) a spacing and twice its value were also found. 
However, those cases are more remarkable (stars No. 78, 92 and 96), where both of the two
most popular spacings were found. They argue against the simplest explanation, namely that 
the sequences represent the consecutive radial order with the same $l$ value. 

The visual inspection is not the fastest way for searching for regular spacing in a large sample. We developed an algorithmic search using the constraints that we learned in the visual inspection as a first trial on the long way to disentangling the pulsation and rotations in $\delta$ Scuti stars. Following this concept we could test that the sequence search algorithm (SSA) properly works. Any extension could come only after the positive test of the first trial.

\subsection{The Algorithm (SSA)}

We present here the Sequence Search Algorithm (SSA) developed for treatment of an even larger sample than ours.
We define the $i$th {\it frequency sequence}
for a given star with $n$ element by
the following set: $S_i=\{ f^{(1)}, f^{(2)}, f^{(3)}, \dots, f^{(n)} \}$,
where $i$ and $n$ are positive integers ($i, n \in \mathbb{N}$). The $S_i$ set is
ordered  $\{ f^{(1)} < f^{(2)} < \dots <f^{(n)} \}$ and
\begin{equation}\label{ser_def}
f^{(j)}+kD-\Delta f \le f^{(j+1)} \le f^{(j)}+kD+\Delta f
\end{equation}
is true for each ($f^{(j)}, f^{(j+1)}$) pair, $j \in \mathbb{N}$, $k=1$ or $k=2$.
$D$ means the {\it spacing}, $\Delta f$ is the {\it tolerance value}.
The upper frequency indices indicate serial numbers within the found
sequence. We define independent lower frequency indices as well
which show the position in the frequency list ordered by decreasing
amplitude vis.
$A(f_1) > A(f_2) > A(f_3), \dots,$.

Since we do not have definite 
knowledge that all modes are excited above an amplitude limit,
we allowed ``gaps'' in the sequences. This means that
the sequence's definition inequality Eq.~(\ref{ser_def}) is fulfilled
for some $j$ indices at $k=2$. Formulating this in another way,
$S_i=\{ f^{(1)}, f^{(2)}, \dots, f^{(j)}, \emptyset, f^{(j+1)},  \dots, f^{(n)} \}$
is considered as a sequence, where $\emptyset$ means the empty set.
We also allow more than one gaps in a sequence, but two subsequent
gaps are forbidden.

{\LongTables
\begin{deluxetable*}{rrrrrrrrrrrrrr}
\tablecaption{List of our sample \label{bigtable}}
\tablehead{
\colhead{No} & \colhead{CoRoT ID} & 
\colhead{$T_{\mathrm {eff}}$} & \colhead{$\log g$} & \colhead{$v_{\mathrm {rad}}$} & 
\colhead{SSF} & \colhead{FF} & \colhead{EF$_{\mathrm {VI}}$} 
& \colhead{SN$_{\mathrm {VI}}$} & \colhead{SP$_{\mathrm {VI}}$} & 
\colhead{EF$_{\mathrm {A}}$} & \colhead{SN$_{\mathrm {A}}$} & \colhead{SP$_{\mathrm {A}}$} & 
\colhead{SP$_{\mathrm {FT}}$} \\
\colhead{}  & \colhead{} & \colhead{(K)} &  \colhead{} & 
\colhead{(km~s$^{-1}$)} & \colhead{} & \colhead{} & \colhead{} & 
\colhead{} & \colhead{(d$^{-1}$)} & \colhead{} & \colhead{} & \colhead{(d$^{-1}$)} & \colhead{(d$^{-1}$)} 
}
\startdata
1=55 & 102661211 & 7075 & 3.575 & 45.0 & 163 & 52 & 25 & 6 & 2.251 & 28,29 & 6,5 & 2.092,1.510 & 0.886 \\
2=66 & 102671284 & 8550 & 3.650 & 87.5 & 130 & 19 & 8 & 1 & 2.137 & 5 & 1 & 2.161 & 2.137 \\
3 & 102702314 & 7000 & 2.975 & 95.0 &141 & 25 & 12 & 3 & 2.169  & 10 & 2 & 2.046 & 0.933  \\
4 & 102712421 & 7400 & 3.950 & 32.5 & 103 & 25 & 13 & 2 & 2.362 & 11 & 2 & 2.356 & 2.294  \\
5 & 102723128 & 6975 & 3.900 & 2.5 & 72 & 18 & 7 & 2 & 1.798 & 8 & 2 & 1.668 & 1.852  \\
6 & 102703251 & 9100 & 3.800 & 42.5 & 118 & 27 & 6 & 1 & 1.850 & 15 & 3 & 1.767 & 1.866  \\
7 & 102704304 & 7050 & 3.250 & 55.0 & 184 & 53 & $-$ & $-$ &   $-$   & 30 & 6 & 1.795 & 0.779  \\
8=92 & 102694610 & 8000 & 3.700 & 55.0 & 193 & 55 & 14 & 3 & 2.470 & 35 & 8 & 2.481 & 4.237  \\
9 & 102706800 & 7125 & 3.325 & 52.5 & 122 & 49 & 27 & 5 & 2.758  & 21,22 & 4,5 & 2.784,3.506 & 1.786  \\
10 & 102637079 & 7325 & 3.850 & 35.0 & 162 & 43 & 21 & 3 & 2.629 & 21 & 4 & 2.614 & 1.374  \\
11=81 & 102687709 & 7950 & 4.400 & 47.5 & 107 & 19 & 9 & 2 & 3.481 & 5 & 1 & 3.570 & 3.472  \\
12 & 102710813 & 8350 & 4.150 & 70.0 &  94 & 13 & 4 & 1 & 2.573 & 4 & 1 & 2.569 & 3.125  \\
13=74 & 102678628 & 7100 & 3.225 & 45.0 & 230 & 49 & $-$ & $-$ & $-$ & 16 & 4 & 2.674 & 2.809  \\
14=96 & 102599598 & 7600 & 4.000 & 65.0 & 99 & 18 & 4 & 1 & 1.844 & 5 & 1 & 1.866 & 3.472  \\
15 & 102600012 & 8000 & 4.400 & 12.5 &  107 & 27 & 9 & 1 & 2.475 & 4,4 & 1,1 & 7.342,2.438 & 2.809  \\
18 & 102618519 & 7500 & 4.500 & 35.0 & 102 & 54 & 10 & 1 & 2.362 & 18,11,16 & 4,2,3 & 6.001,2.359,3.345 & 2.232 \\
19 & 102580193 & 7525 & 4.150 & 50.0 & 125 & 43 & 7 & 1 & 3.531  & 8,8 & 2,2 & 6.175,4.023 & 3.205  \\
20 & 102620865 & 11250 & 3.975 & 50.0 & 244 & 40 & $-$ & $-$ & $-$ & 19 & 4 & 1.974 & 1.097  \\
21 & 102721716 & 7700 & 4.150 & 25.0 & 149 & 52 & 21 & 3 & 2.537 & 9,5  & 2,1 & 7.492,2.636 & 2.427 \\
22 & 102622725 & 6000 & 4.300 & $-$20 & 144 & 23 & 15 & 4 & 3.497 & 5,5 & 1,1 & 1.877,2.598 & 4.464  \\
23 & 102723199 & 6225 & 3.225 & 40.0 & 113 & 22 & 9 & 3 & 3.364  & 10 & 2 & 1.461 & 1.316  \\
24 & 102623864 & 7900 & 4.000 & 50.0 & 117 & 30 & 16 & 4 & 2.226 & 8 & 2 & 3.320 & 2.294  \\
25 & 102624107 & 8400 & 4.050 & 57.5 & 70 & 32 & 4 & 1 & 3.215 & 8 & 2 & 3.299 & 2.100  \\
26 & 102724195 & 7550 & 3.900 & 42.5 & 58 & 28 & 14 & 3 & 3.362 & 10,9 & 2,2 & 3.200,2.728 & 1.208  \\
27 & 102728240 & 7450 & 4.200 & 25.0 & 168 & 55 & 20 & 4 & 3.255 & 18,18 & 4,4 & 5.995,3.178 & 1.623  \\
28 & 102702932 & 6975 & 3.350 & 47.5 & 155 & 48 & 16 & 4 & 3.247 & 26 & 6 & 2.655 & 0.806  \\
29 & 102603176 & 12800 & 4.300 & 35.0 & 308 & 64 & 29 & 6 & 2.342 & 35 & 7 & 2.389 & 0.984  \\
30 & 102733521 & 7125 & 3.625 & 50.0 & 174 & 43 & 18 & 3 & 3.267 & 16,17 & 4,3 & 3.437,2.297 & 1.667  \\
31 & 102634888 & 7175 & 4.000 & 40.0 & 179 & 39 & $-$ & $-$ & $-$ & 15 & 3 &  2.622 & 1.344  \\
32 & 102735992 & 7225 & 3.800 & 62.5 & 83 & 38 & 10 & 2 & 3.117 & 16 & 4 & 3.253 & 1.552  \\
33 & 102636829 & 7000& 3.200 & 42.5 &  93 & 43 & 9 & 2 & 2.303  & 21,23 & 5,5 & 2.396,1.540 & 1.282  \\
34 & 102639464 & 9450 & 3.900 & 52.5 & 141 & 31 & $-$ & $-$ & $-$ & 5 & 1 & 3.099 & 3.333  \\
35 & 102639650 & 7500 & 3.900 & 32.5 & 78 & 28 & 16 & 3 & 3.484 & 8,9 & 2,2 & 3.492,2.609 & 3.387  \\
36 & 102641760 & 7950 & 4.300 & 40.0 & 135 & 32 & $-$ & $-$ & $-$ & 9 & 2 & 2.723 & 2.632  \\
37 & 102642516 & 7275 & 3.700 & 45.0 & 72 & 20 & 8 & 2 & 2.335 & 5 & 1 & 2.586 & 3.012  \\
38 & 102742700 & 7550 & 3.875 & 15.0 & 121 & 28 & 14 & 3 & 2.443 & 5 & 1 & 2.910 & 2.404  \\
39 & 102743992 & 7950 & 4.300 & 42.5 & 126 & 20 & 6 & 1 & 2.454  & 6 & 1 & 4.382 & 4.386  \\
40 & 102745499 & 7900 & 3.850 & 80.0 & 119 & 22 & 10 & 3 & 2.603 & 8 & 2 & 1.747 & 1.323  \\
43 & 102649349 & 9425 & 3.950 & 65.0 & 121 & 16 & 4 & 1 & 1.949 & 5 & 1 & 1.947 & 2.119  \\
45 & 102647323 & 8200 & 4.300 & 67.5 & 100 & 32 & 7 & 1 & 2.379 & 8,9 & 2,2 & 3.306,1.407 & 3.846  \\
47 & 102650434 & 8500 & 3.875 & 72.5 & 210 & 34 & 13,14 & 3,4 & 1.597,2.525 & 11 & 2 & 1.611 & 1.092  \\
48 & 102651129 & 8350 & 3.750 & 40.0 & 88 & 35 & 12 & 2 & 3.413 & 13 & 3 & 3.464 & 3.521  \\
49 & 102753236 & 7600 & 4.100 & 32.5 & 375 & 37 & 12 & 2 & 3.767 & 14 & 3 & 2.317 & 2.604  \\
50 & 102655408 & 7375 & 4.000 & 42.5 & 75 & 28 & 14,6 & 3,1 & 3.394,1.550 & 8 & 2 & 3.936 & 2.747  \\
51 & 102655654 & 7200 & 3.675 & 72.5 & 97 & 16 & 4 & 1 & 3.377 & 4 & 1 & 1.867 & 3.378  \\
52 & 102656251 & 7950 & 4.200 & 60.0 & 128 & 22 & 4 & 1 & 3.288 & 10 & 2 & 2.747 & 1.623  \\
53 & 102657423 & 8150 & 3.425 & 52.5 &  161 & 36 & 10 & 2 & 2.523 & 18 & 4 & 2.492 & 2.403  \\
54 & 102575808 & 7250 & 3.325 & 17.5 & 202 & 47 & 22,37 & 4,6 & 4.659,2.289 & 17,18 & 4,4 & 2.300,3.275 & 4.717 \\
55=1 & 102661211 & 7075 & 3.575 & 45.0 & 163 & 43 & 9 & 3 & 2.337 & 21,24 & 5,5 & 2.544,2.262 & 0.874  \\
56 & 102761878 & 7375 & 3.700 & 32.5 & 80 & 11 & 4 & 1 & 2.564 & $-$ & $-$ & $-$ &  4.310  \\
62 & 102576929 & 8925 & 4.050 & 32.5 & 104 & 20 & 7 & 2 & 6.365 & 9 & 2 & 1.834 & 1.748  \\
63 & 102669422 & 7300 & 3.675 & 50.0 & 82 & 35 & 14 & 2 & 3.390  & 18 & 4 & 3.285 & 1.712  \\
65 & 102670461 & 7325 & 3.575 & 50.0 & 142 & 49 & 22 & 4 & 3.459 & 21 & 4 & 3.437 & 1.282  \\
66=2 & 102671284 & 8550 & 3.650 & 87.5 & 130 & 39 & 10 & 2 & 2.152 & 16 & 4 & 2.406 & 2.119  \\
67 & 102607188 & 8100 & 4.200 & 40.0 & 95 & 23 & $-$ & $-$ & $-$ & 4 & 1 & 3.101 & 3.425  \\
68 & 102673795 & 8050 & 3.750 & 27.5 & 65 & 13 & $-$ & $-$ & $-$ & 5 & 1 & 1.929 & 2.119  \\
69 & 102773976 & 7525 & 4.400 & 17.5 & 52 & 13 & $-$ & $-$ & $-$ & 4 & 1 & 4.682 & 3.731  \\
70 & 102775243 & 7950 & 4.250 & 50.0 & 126 & 31 & 10,4 & 2,1 & 4.167,3.002 & 8 & 2 & 3.059 & 3.676  \\
71 & 102775698 & 9550 & 3.750 & 22.5 & 473 & 56 & 24 & 4 & 3.351 & 30,28  & 6,6 & 3.277,2.218 & 1.131  \\
72 & 102675756 & 7350 & 3.175 & 77.5 & 342 & 40 & 23 & 4 & 2.277 & 23,25 & 5,5 & 2.249,1.977 & 2.137  \\
73 & 102677987 & 7700 & 3.950 & 37.5 & 102 & 26 & 13 & 3 & 3.293 & 8,10 & 2,2 & 3.416,2.417 & 1.176  \\
74=13 & 102678628 & 7100 & 3.225 & 20.0 & 230 & 68 & 32 & 6 & 3.343 & 37 & 8 & 2.940 & 0.647  \\
75 & 102584233 & 6400 & 3.725 & 75.0 & 58 & 12 & 6 & 2 & 3.287 & $-$ & $-$ & $-$ & 3.472  \\
76 & 102785246 & 7425 & 3.800 & 30.0 & 76 & 37 & 20 & 5 & 3.527 & 21,21 & 4,4 & 1.772,2.067 & 1.761  \\
77 & 102686153 & 7125 & 3.525 & 45.0 & 106 & 31 & 10,19 & 2,6 & 2.867,5.713 & 9,9 & 2,2 & 2.521,3.692 & 2.033  \\
78 & 102786753 & 7100 & 3.425 & 55.0 & 238 & 59 & 22,11 & 4,2 & 2.543,3.297 & 29 & 6 & 2.392 & 1.101  \\
79 & 102787451 & 7300 & 4.000 & 37.5 & 76 & 13 & 6 & 2 & 3.428 & 4 & 1 & 3.357 & 3.676  \\
80 & 102587554 & 7375 & 3.700 & 47.5 &  82 & 34 & 13,14 & 3,2 & 4.293,2.487 & 11,15,12 & 2,3,3 & 4.247,1.734,3.365 & 1.712  \\
81=11 & 102687709 & 7950 & 4.400 & 47.5 & 107 & 36 & $-$ & $-$ & $-$ & 8 & 2 & 3.480 & 4.032  \\
82 & 102688156 & 7725 & 4.400 & 55.0 & 96 & 21 & 7 & 1 & 2.308 & 5 & 1 & 4.098 & 4.032  \\
83 & 102788412 & 8000 & 3.925 & 70.0 & 47 & 10 & 5 & 1 & 2.357 & $-$ & $-$ & $-$ & 6.250  \\
84 & 102688713 & 7300 & 4.150 & 47.5 & 111 & 40 & 4 & 1 & 3.584 & 17 & 4 & 2.699 & 2.500  \\
86 & 102589546 & 7250 & 3.700 & 27.5 & 178 & 35 & 17 & 3 & 2.599 & 13,12 & 3,2 & 4.890,2.591 & 2.551  \\
87 & 102690176 & 7425 & 3.525 & 60.0 & 111 & 35 & 20 & 4 & 2.551 & 17 & 4 & 1.458 & 4.386  \\
88 & 102790482 & 7225 & 3.475 & 52.5 & 125 & 48 & 15 & 3 & 2.704 & 19 & 4 & 2.837 & 2.358  \\
89 & 102591062 & 7600 & 3.650 & 30.0 & 101 & 10 & 6 & 1 & 2.551 & $-$ & $-$ & $-$ & 6.944  \\
90 & 102691322 & 7650 & 4.050 & 37.5 & 45 & 18 & $-$ & $-$ & $-$ & 4,4 & 1,1 & 7.170,3.645 & 3.497  \\
91 & 102691789 & 7800 & 3.750 & 75.0 & 58 & 20 & 9 & 2 & 2.648 & 5 & 1 & 2.803 & 6.250  \\
92=8 & 102694610 & 8000 & 3.700 & 55.0 & 193 & 53 & 30,22 & 5,5 & 2.454,3.471 & 35,38 & 7,7 & 2.576,1.880 & 4.032  \\
93 & 102794872 & 7575 & 4.150 & 32.5 & 157 & 58 & 8 & 1 & 4.346 & 20 & 4 & 4.219 & 1.706  \\
94 & 102596121 & 7700 & 4.000 & 22.5 & 92 & 33 & $-$ & $-$ & $-$ & 7 & 1 & 3.445 & 2.564  \\
95 & 102598868 & 7750 & 3.900 & 35.0 & 76 & 26 & 6 & 2 & 3.003 & 10,8 & 2,2 & 2.462,3.294 & 2.564  \\
96 & 102599598 & 7600 & 4.000 & 65.0 & 99 & 55 & 22,19 & 5,4 & 2.429,3.387 & 42,37 & 9,7 & 2.584,1.835 & 1.552 
\enddata
\tablecomments{
Columns:
(1) the running number (No.),
(2) the CoRoT ID,
(3) the effective temperature ($T_{\mathrm{eff}}$),
(4) the surface gravity ($\log g$),
(5) the radial velocity ($v_{\mathrm {rad}}$),
(6) the number of SigSpec frequencies (SSF),
(7) the number of filtered frequencies (FF),
(8) the number of frequencies included in the sequences from the VI (EF$_{\mathrm {VI}}$),
(9) the number of sequences from the VI (SN$_{\mathrm {VI}}$),
(10) the dominant spacing from the VI (SP$_{\mathrm {VI}}$),
(11) the number of frequencies included in the sequences from the SSA (EF$_{\mathrm {A}}$),
(12) the number of sequences from the SSA (SN$_{\mathrm {A}}$),
(13) the dominant spacing from the SSA (SP$_{\mathrm {A}}$),
(14) the spacing from the FT (SP$_{\mathrm {FT}}$).
}
\end{deluxetable*} 
}

\begin{figure*}
\includegraphics[width=15.5cm]{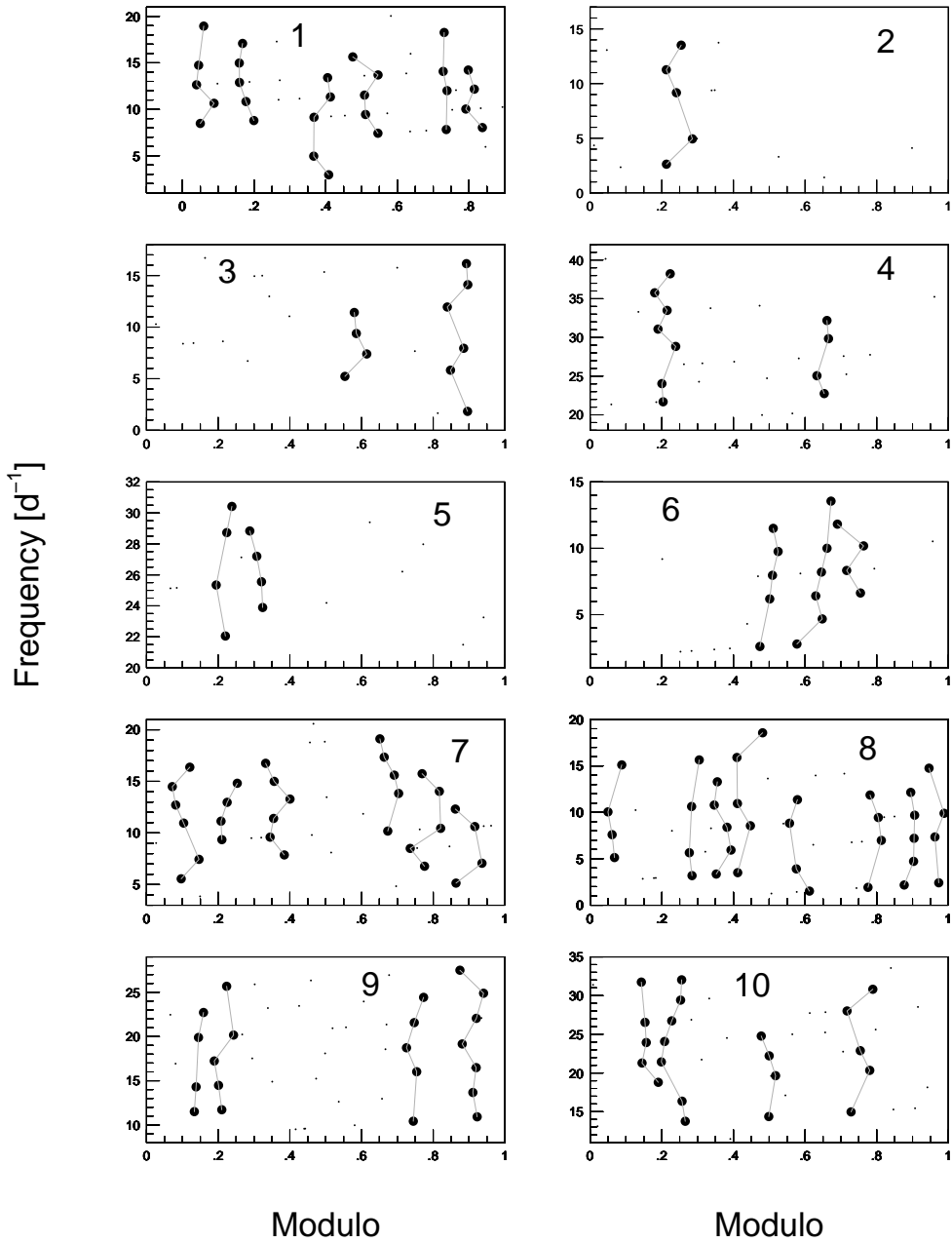}
\caption[]{
Echelle diagrams using the best spacing obtained by SSA. The labels mark the running 
number of stars in our sample. The spacings used for modulo calculation 
are 2.092, 2.161, 2.046, 2.356, 1.668, 1.767, 1.795, 2.481, 2.784, and 2.614 d$^{-1}$ for 
the increasing running numbers, respectively.
} \label{fig3}
\end{figure*}

\begin{figure*}
\includegraphics[width=15.5cm]{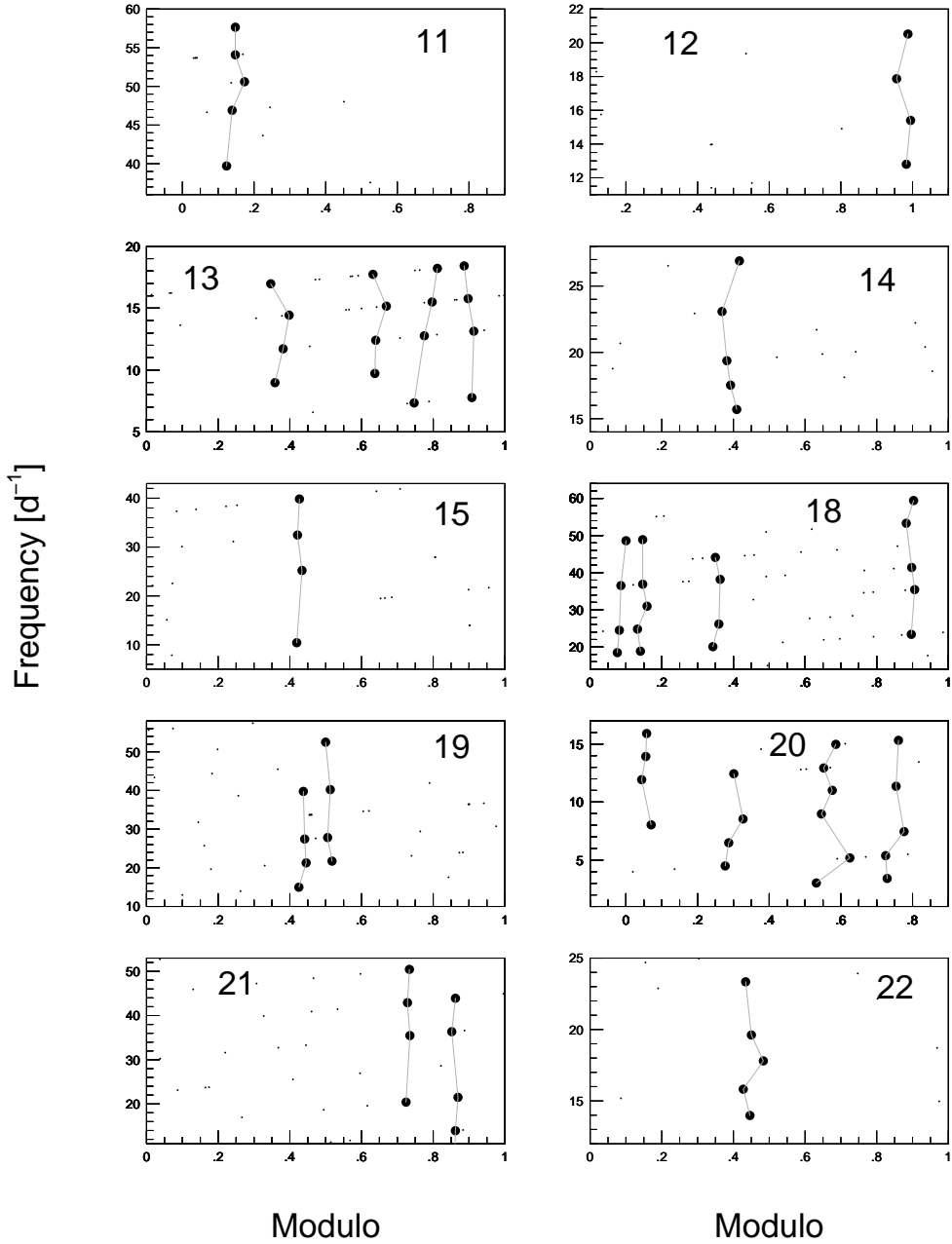}
\caption[]{
Echelle diagrams using the best spacing obtained by SSA. The labels mark the running number of 
stars in our sample. The spacings used for modulo calculation 
are 3.570, 2.569, 2.674, 1.866, 7.342, 6.001, 6.175, 1.478, 7.492, and 1.877 d$^{-1}$ for 
the increasing running numbers, respectively.
} \label{fig4}
\end{figure*}

\begin{figure*}
\includegraphics[width=15.5cm]{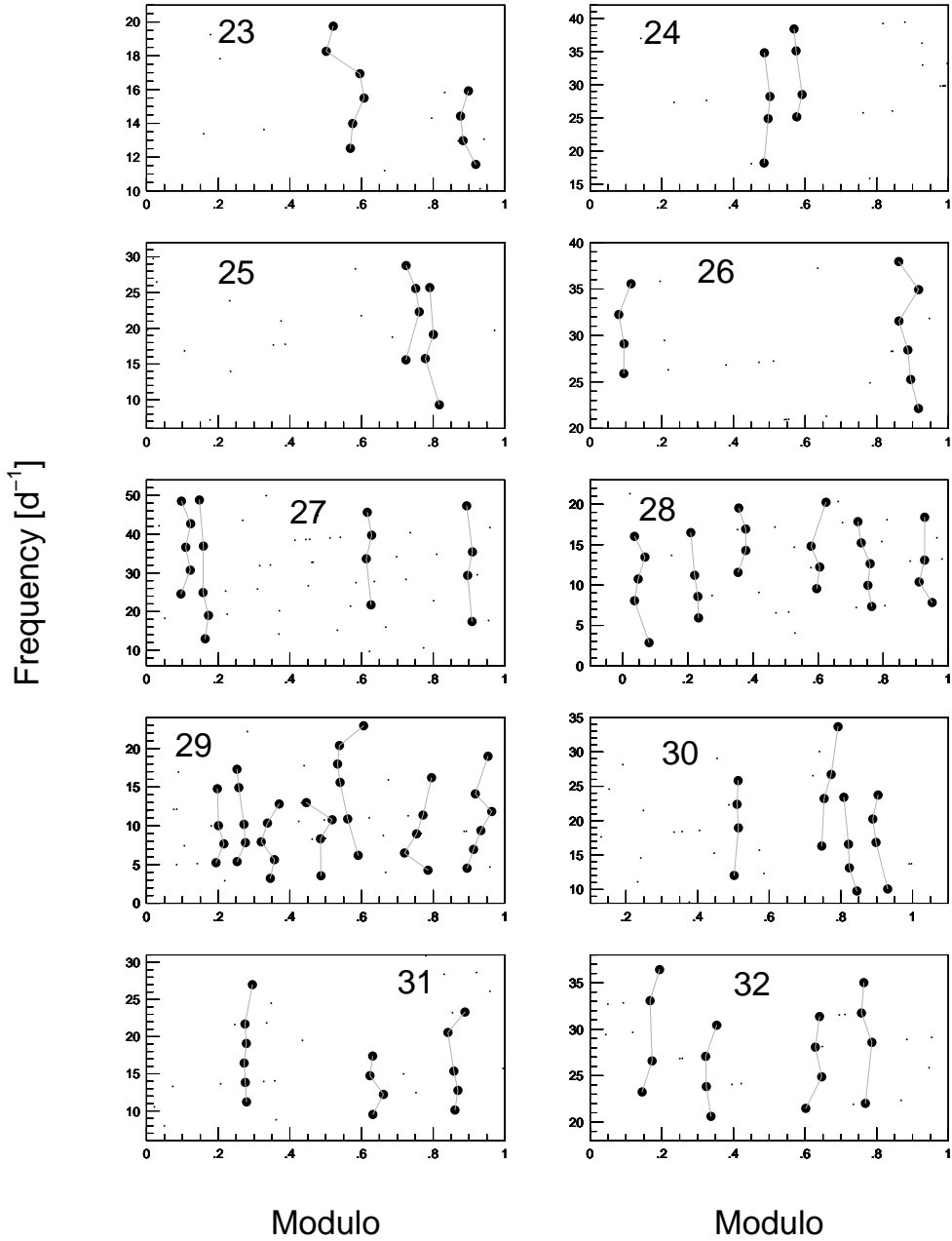}
\caption[]{
Echelle diagrams using the best spacing obtained by SSA. The labels mark the running number of stars in our sample. 
The spacings used for modulo calculation are 1.461, 3.320, 3.299, 3.200, 5.995, 2.655, 2.389, 3.082, 2.622, 
and 1.671 d$^{-1}$ for the increasing running numbers, respectively.
} \label{fig5}
\end{figure*}

\begin{figure*}
\includegraphics[width=15.5cm]{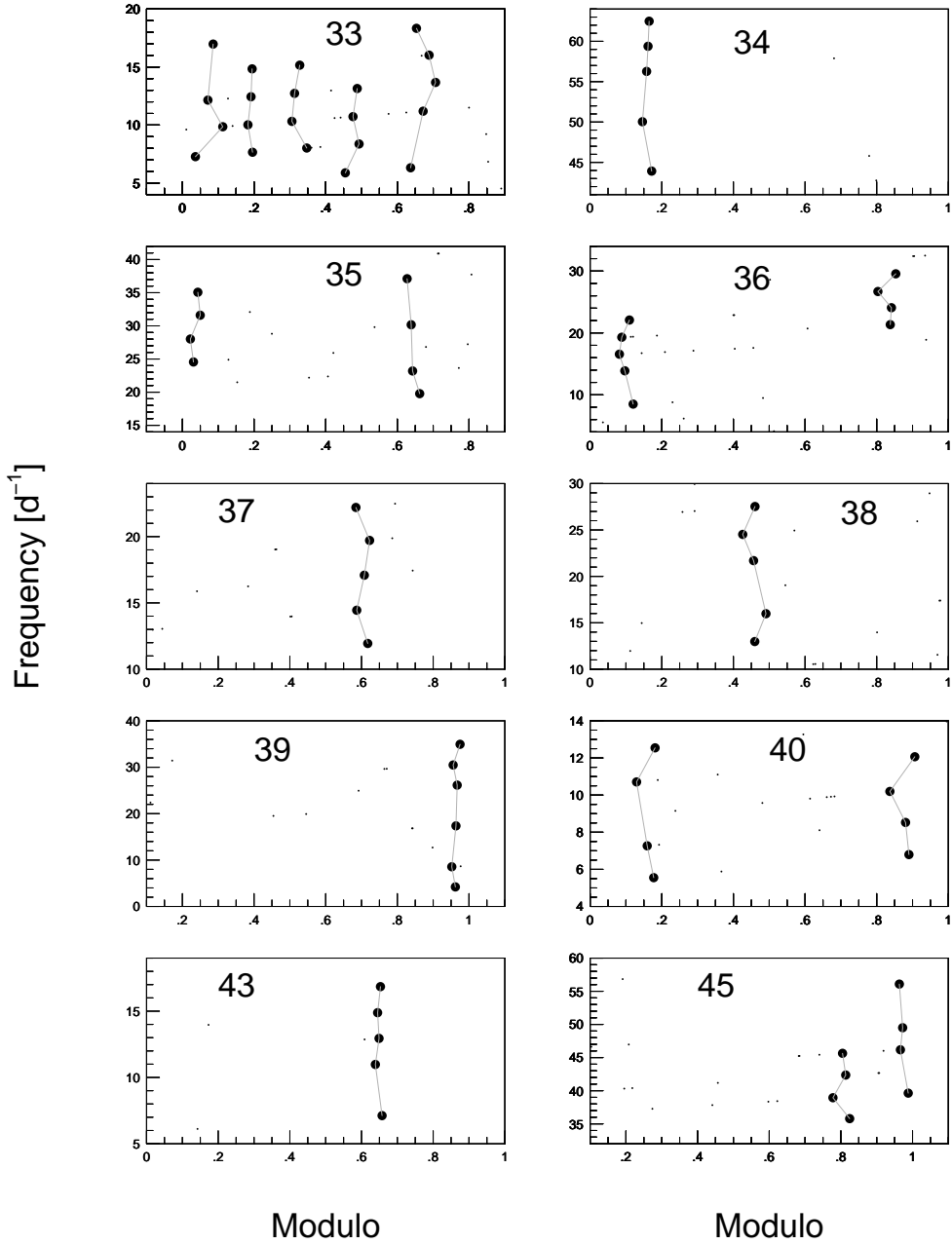}
\caption[]{
Echelle diagrams using the best spacing obtained by SSA. The labels mark the running 
number of stars in our sample. The spacings used for modulo calculation are 
2.396, 3.099, 3.492, 2.723, 2.586, 2.910, 4.382, 1.747, 1.947, and 3.306 d$^{-1}$ for 
the increasing running numbers, respectively.
} \label{fig6}
\end{figure*} 

\begin{figure*}
\includegraphics[width=15.5cm]{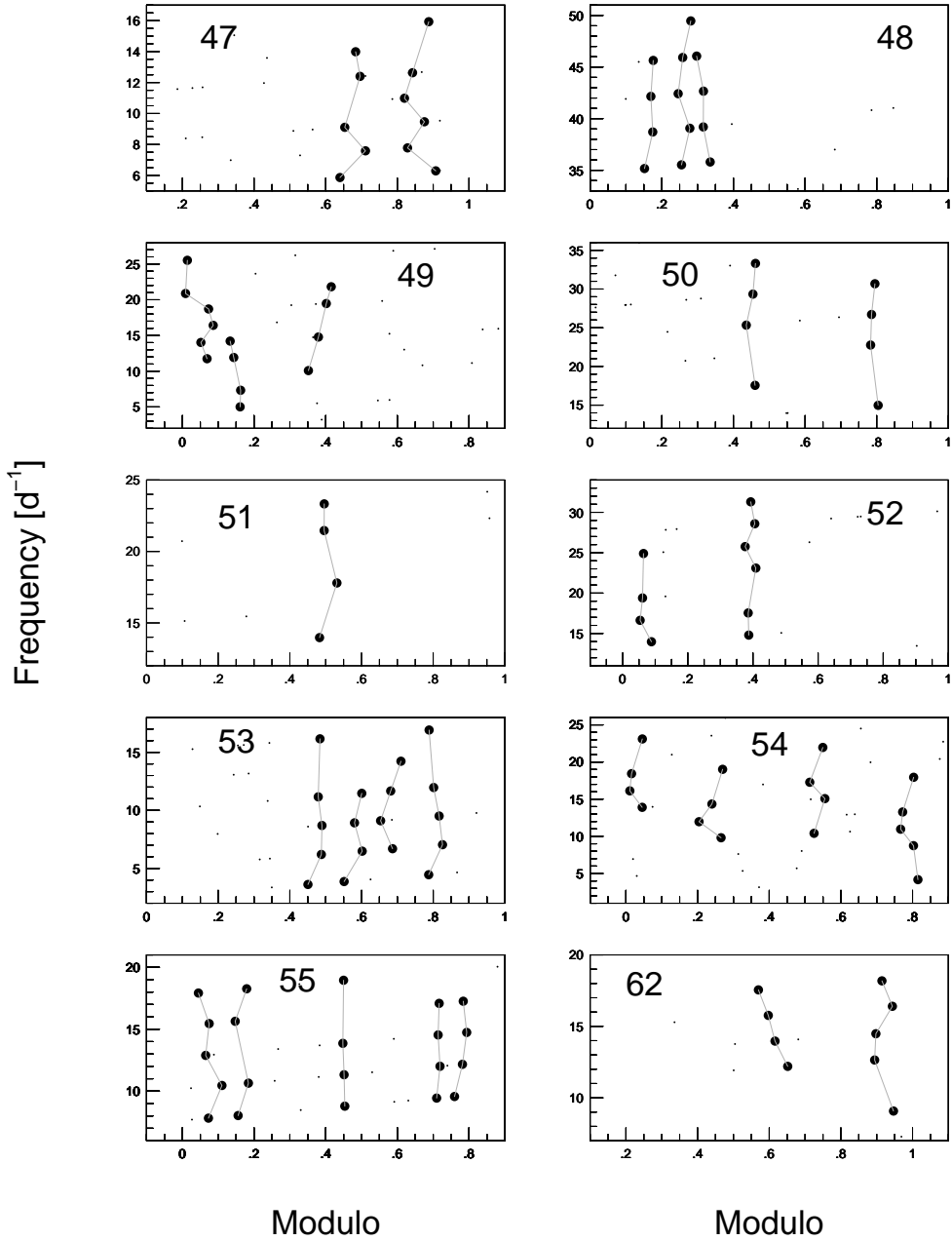}
\caption[]{
Echelle diagrams using the best spacing obtained by SSA. The labels mark the 
running number of stars in our sample. The spacings used for modulo calculation are 
1.611, 3.464, 2.317, 3.936, 1.867, 2.748, 2.492, 2.300, 2.544, and 1.834 d$^{-1}$ for 
the increasing running numbers, respectively.
} \label{fig7}
\end{figure*}

\begin{figure*}
\includegraphics[width=15.5cm]{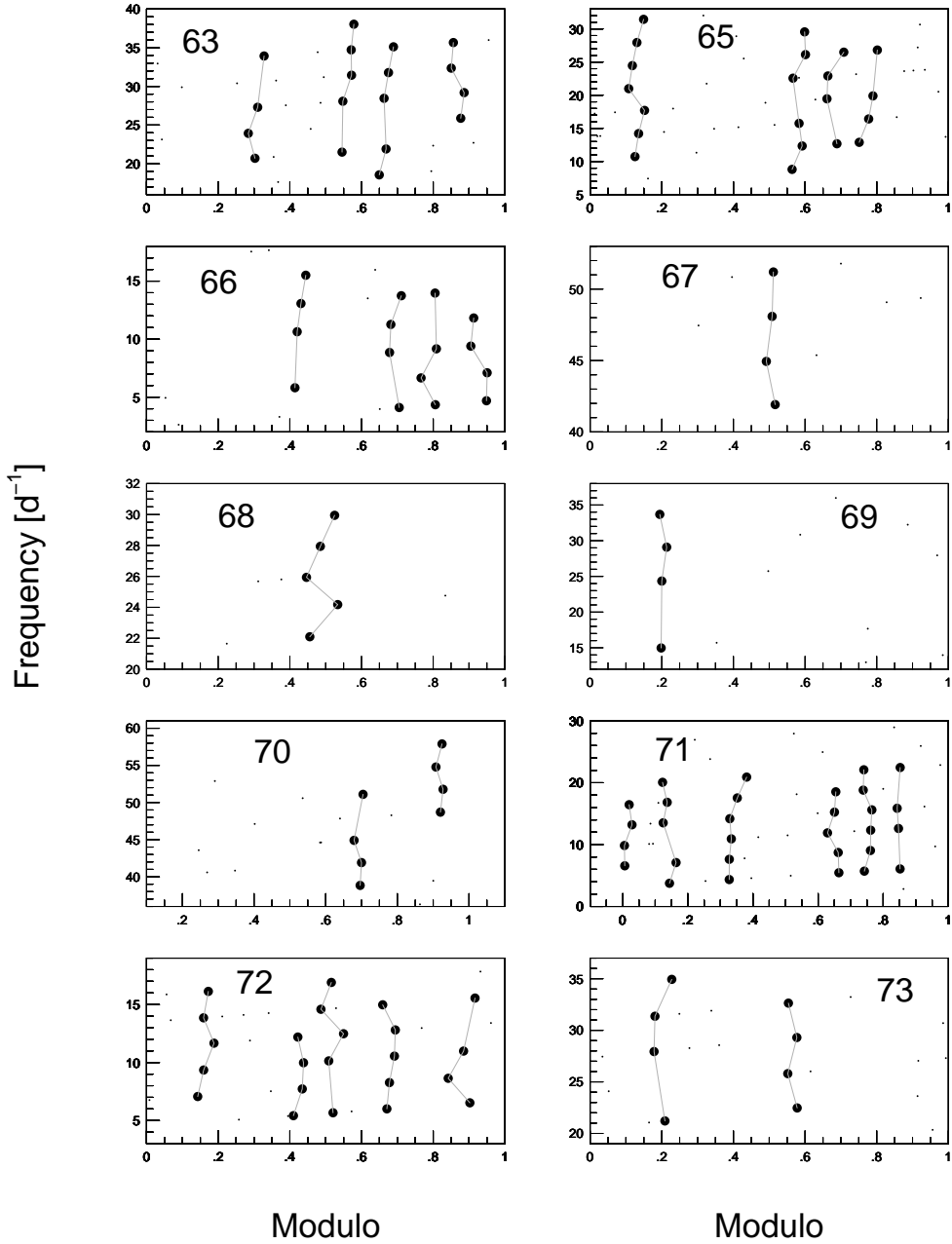}
\caption[]{
Echelle diagrams using the best spacing obtained by SSA. 
The labels mark the running number of stars in our sample. The spacings used for modulo 
calculation are 3.285, 3.437, 2.406, 3.101, 1.929, 4.682, 3.059, 3.495, 2.249, and 3.416 d$^{-1}$ 
for the increasing running numbers, respectively.
} \label{fig8}
\end{figure*}

\begin{figure*}
\includegraphics[width=15.5cm]{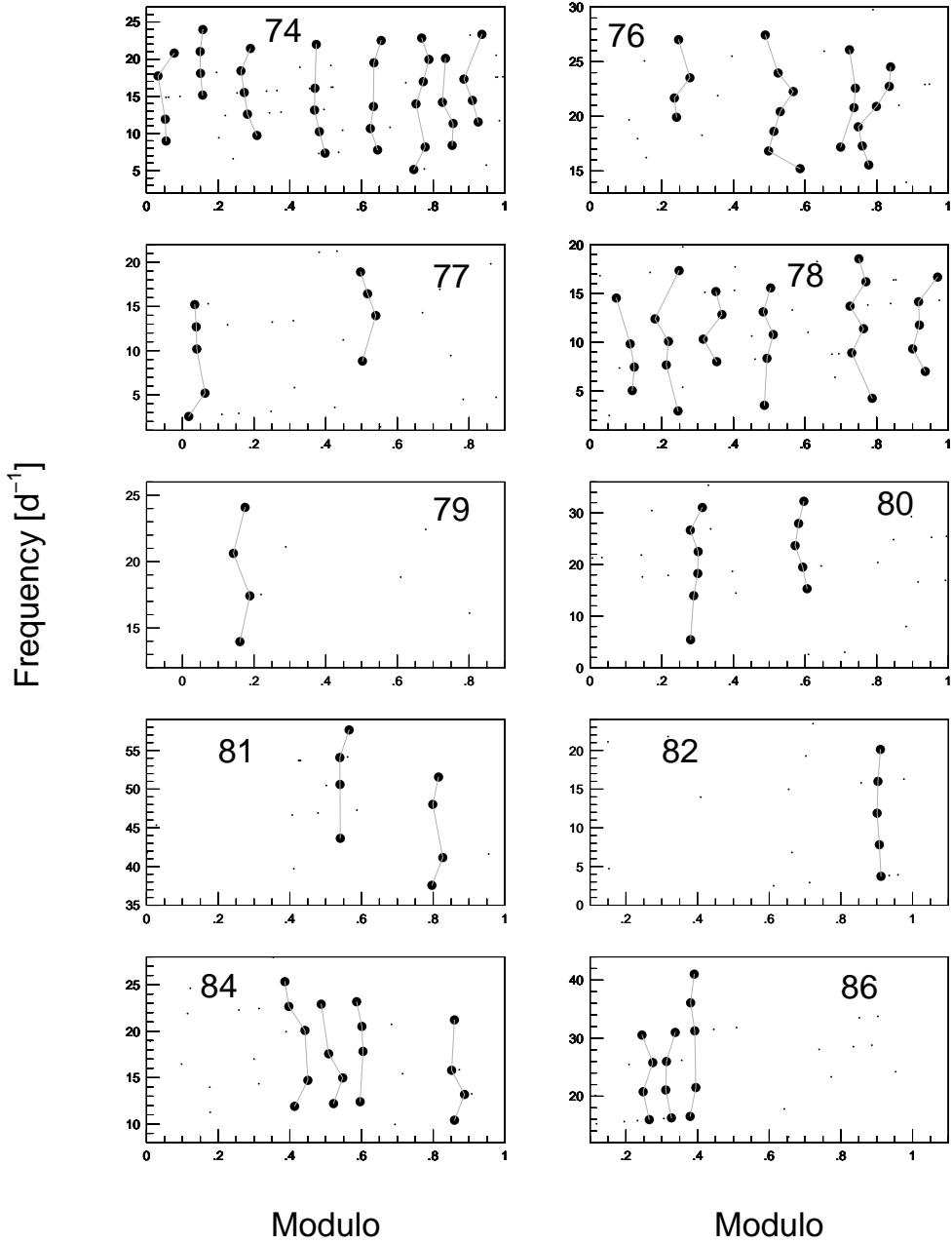}
\caption[]{
Echelle diagrams using the best spacing obtained by SSA. The labels mark the running number of stars 
in our sample. The spacings used for modulo calculation are 2.940, 1.772, 2.521, 2.392, 3.357, 
4.247, 3.480, 4.098, 2.699, and 4.890 d$^{-1}$ for the increasing running numbers, respectively.
} \label{fig9}
\end{figure*}

\begin{figure*}
\includegraphics[width=15.5cm]{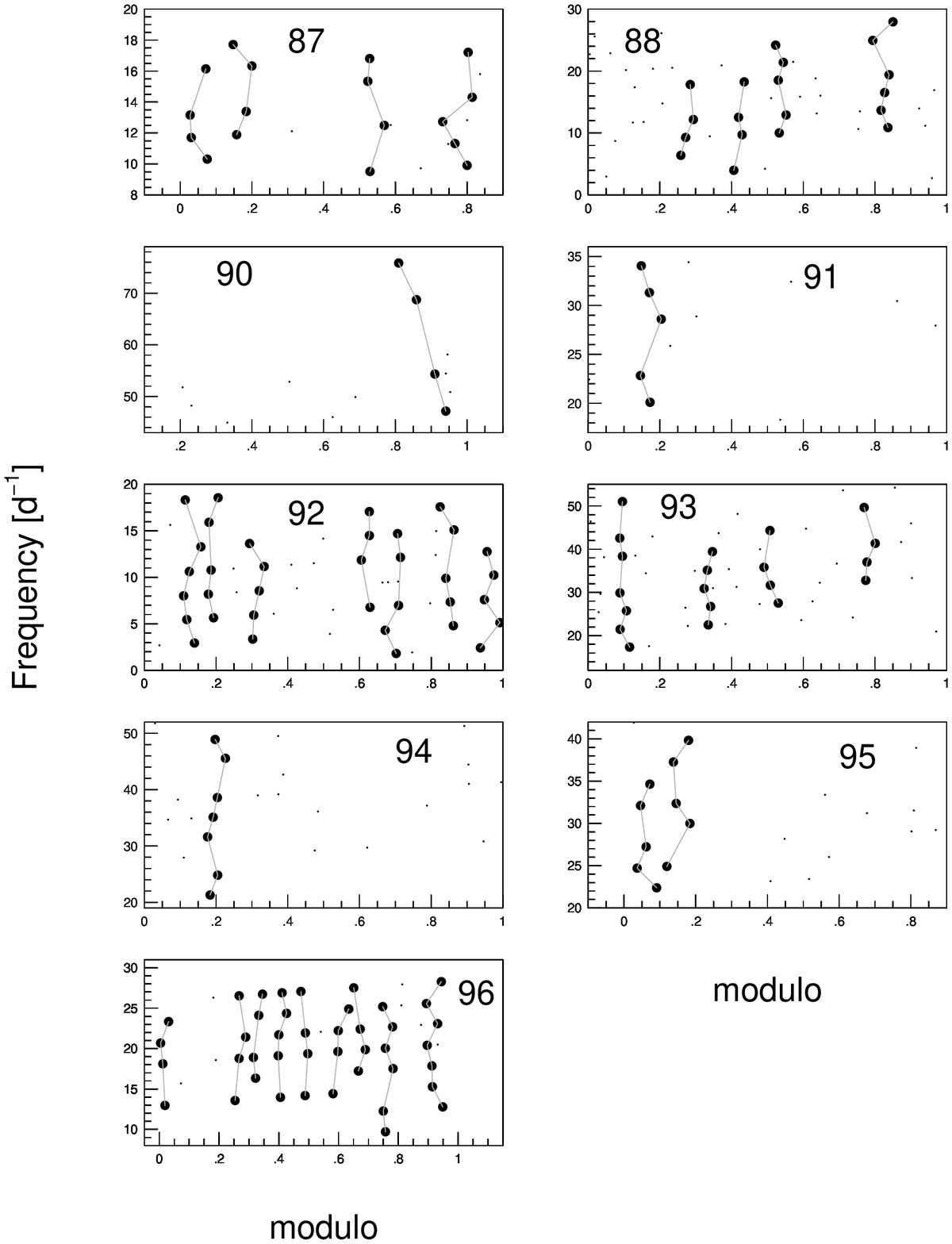}
\caption[]{
Echelle diagrams using the best spacing obtained by SSA. The labels mark the running number 
of stars in our sample. The spacings used for modulo calculation are 1.862, 2.837, 7.170, 
2.803, 2.576, 4.219, 3.445, 2.462, and 2.464 d$^{-1}$ for the increasing running numbers, respectively.
} \label{fig10}
\end{figure*}

The SSA scans through the frequency lists and selects frequency
sequences defined by Eq.~(\ref{ser_def}) with a parameter set $D$, $\Delta f$ and $n$.
The search begins from the highest amplitude frequency $f_1$ that we
called {\it basis frequency}. The search proceeds with the frequency
$\hat{f}_1$ the closest neighbor of $f_1$,
if $\vert \hat{f}_1- f_1 \vert \le D$ and
$\vert \hat{f}_1- f_1 \vert > \Delta f$. If the $\hat{f}_1$ is too close (viz.
$\vert \hat{f}_1- f_1 \vert \le \Delta f$), the algorithm steps to the next frequency
 $\hat{f}_2$ and so on.
 We collect the sequences $S_1, S_2, \dots, S_N$, ($N \le i$) found by the search from
the frequencies $f_1$, $\hat{f}_1$, $\hat{f}_2$, $\dots$,$\hat{f}_{i-1}$
as a {\it pattern} belonging to
a given $D$ and basis frequency $f_1$. Next, the algorithm goes to
the second highest-amplitude frequency $f_2$ and (if it is not the
element of the previous pattern) begins to collect a new
pattern again. On the basis of the VI (Sec.~\ref{visual})
we demanded that at least one of the two highest amplitude frequencies
must be in a pattern, so we did not search from smaller amplitude frequencies
($f_i$, $i \ge3$) as a basis frequency.

Starting from the parameter range obtained by the VI
we made numerical experiments determining the optimal input parameters.
We found the smallest difference between the results of
the automatic and visual sequence search at $\Delta f=0.1$~d$^{-1}$.
Since we do not have any other reference point, we fixed $\Delta f$ at the this value.
If we chose $n$ (the length of the sequence)
to be small ($n \le 3$) we obtained a huge number of short sequences for
most of the stars. Avoiding this, we set $n=4$.
The crucial parameter of the algorithm is the spacing $D$.
Our program determines $D$ in parallel
with the sequences. The primary searching interval was
$D_{\mathrm {min}} = 1.5 \le D \le 7.8 = D_{\mathrm {max}}$. The lower limit was fixed according to our 
results obtained by the VI. To reduce the computation time we applied an adaptive grid
instead of an equidistant one.
We calculated the spacings between the ten highest
amplitude frequencies for each star $D_{1,2}=\vert f_1 - f_2 \vert$,
$D_{1,3}=\vert f_1 - f_3 \vert$, $\dots$, $D_{9,10}=\vert f_9 - f_{10} \vert$.
The $D_{l,m}$ values could be either too high or too low for a large separation,
therefore we selected those ones where $D_{\mathrm {min}} \le D_{l,m} \le D_{\mathrm {max}}$
and restricted our further investigations to this selected  $D_{l,m}$.
 Then we define a fine grid around all
such spacings with $D_{l,m,h}=D_{l,m}\pm h \delta f$,
where $h=15$ and $\delta f = 0.01$~d$^{-1}$. The SSA script ran for all
$D=D_{l,m,h}$ searching for possible sequences for all $D$ values.

The SSA script calculates (1) the total number of frequencies in all series
for a given $D$, which is the frequency number of the pattern,
(2) the number of found sequences, (3) the actual standard deviation of the
echelle ridges and (4) the amplitude sum of the pattern frequencies.
These four output values helped us to recognize the dominant spacing, since the algorithm 
revealed in many stars two or three characteristic spacings.
The similar parameters that we derived by VI, the number of the frequencies 
in the sequences (EF$_{\mathrm A}$), the number of sequences (SN$_{\mathrm A}$) and the spacings 
(SP$_{\mathrm A}$) are given in the 11th, 12th and 13th column 
of Table~\ref{bigtable}. The algorithmic search recognized many more spacing values. 
Obviously, when we have more spacings, the appropriate set of frequencies and the number 
of frequencies are also given. The best solutions are given at the first place of the columns.

The sequences obtained by SSA are also flagged in the electronic table
in additional columns (see in Table~\ref{sample_data}). SSA1, SSA2..., etc. agree with 
the first, second,..., etc. value of the spacing. 
The flags are similar as in the case of VI ($0$ -- not included, $1, 2, 3,\dots$, are 
the frequencies of the 1st, 2nd, 3rd, $\dots$, sequences).

The summary of the results on SSA is the following. SSA found independent solutions for 73 stars. 
Unexpectedly, the test cases showed seemingly more diversity. 
As we noticed from the beginning, the filtering process resulted, for some cases, in 
quite different number of frequencies used in the SSA. Comparing to the number of the SigSpec frequencies, 
the differences in the resulted frequency content of the double-checked stars is not remarkable, 
in most cases less than 10\%. In any case there are block of frequencies of highest amplitudes that 
are common to both files of the double checked cases. This guarantees that the SSA uses the same basis 
frequencies for the sequence search. Keeping the differences of the frequency content of the 
double-checked cases, we intended to check the sensitivity of the SSA to the frequency content.
It is obvious that if we have a larger frequency content, then we find more sequences and more 
frequencies located on the echelle ridges. Of course, this will also influence the mean spacings. 
Nevertheless, as Table~\ref{bigtable} shows, the spacings differ by less than 10\%.

The comparison of the two approaches, VI and the SSA gives the following result.
They resulted in similar spacing for 42 
stars. In the SSA we found six cases with half of the VI values. In 23 cases
different spacing values were found. The seemingly large number contains the cases where 
we did not find any sequences in the star by one of the two approaches (12 for VI and 4 for SSA. 
There is no overlap in these subsets). 

The best spacings found by the algorithm for 
the CoRoT targets (the first value of 13th column) are used to create the
echelle diagrams presented in Figs.~\ref{fig3}-\ref{fig10}. All filtered frequencies are plotted (small and large dots), while the frequencies located on an echelle ridge are marked by large dots. Taking into account the fixed $\pm$0.1 d$^{-1}$ tolerance we may not expect to find any effects caused by the change in chemical composition (glitches) or effects caused by the evolution (avoided crossing). However, we may conclude that we found unexpectedly large numbers of regular frequency spacing in our sample of CoRoT $\delta$ Scuti stars. Any relation that we find among the echelle ridges, the physical parameters and the estimated rotational splitting confirms that the echelle ridges are not an accidental arrangement of unrelated frequencies along an echelle ridge.

\subsection{Fourier Transform (FT)}\label{FT}

\begin{figure}
\includegraphics[width=9cm]{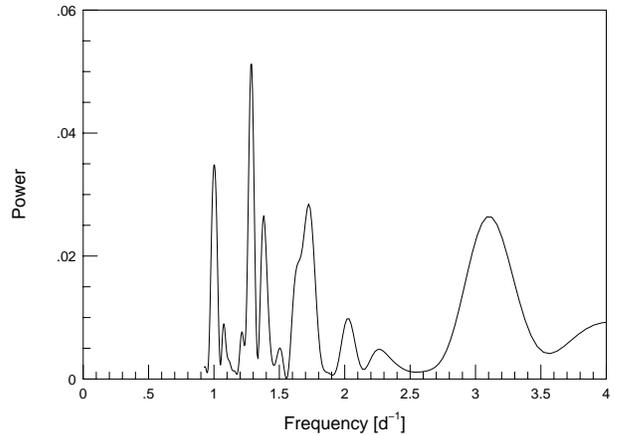}
\caption[]{
Fourier Transform of star No. 65. The highest peak at 1.282 d$^{-1}$ agrees with 
a shift of sequences in VI. The lower amplitude peak agrees with 3.459 or 3.437 d$^{-1}$ 
spacings that are obtained by VI and SSA, respectively.
} \label{fig11}
\end{figure}

\begin{figure}
\includegraphics[width=9cm]{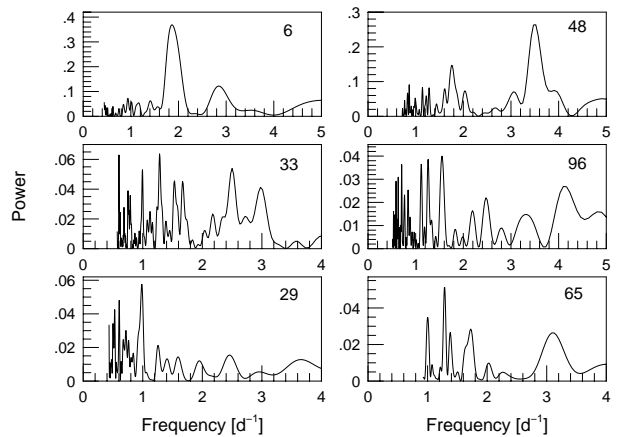}
\caption[]{
Some characteristic examples of FT in our sample. The labels mark the running 
number of the star. The simplest and the most complex examples are in the top and 
the middle panels. The bottom panels show examples with very low value of the spacing. 
Both VI and SSA resulted in higher values. The highest peak probably represents a shift between sequences.
} \label{fig12}
\end{figure}

Fourier Transform (FT) of the frequencies involved in the pulsation is, nowadays, widely used 
in searching period spacing and finding the large separation since \citet{Handler97} to \citet{Garcia15}. It is worthwhile to compare the 
spacing obtained by FT and by our sequence search method. We 
followed the way described by \citet{Handler97} (instead of the way introduced by \citealt{Moya10}) and derived the FT spacing (the highest peak) 
for our sample, given in the 14th column of Table~\ref{bigtable}. 

The FT of the star No. 65 is shown in Fig.~\ref{fig11}. The highest peak suggests a 
large separation at 1.282 d$^{-1}$ that does not agree with the spacing 
obtained by the VI and SSA (3.459 and 3.437 d$^{-1}$, respectively). 
FT spacing is closer to the characteristic shifts derived for the third sequence relative to the first one (1.209 d$^{-1}$) to the leftward 
direction. 
The FT shows a peak near our value but it is definitely not the highest peak. 

\begin{figure}
\includegraphics[width=9cm]{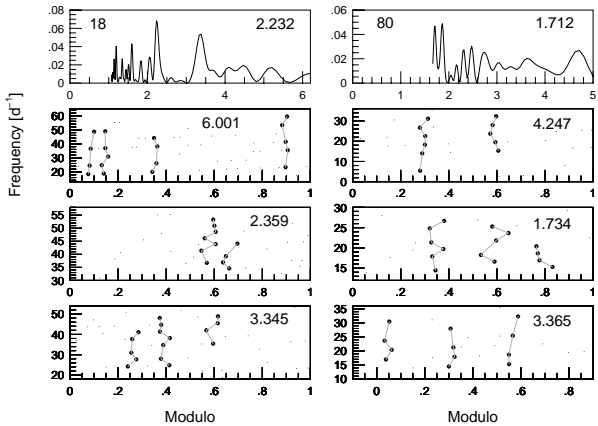}
\caption[]{
Comparison of FT diagram and echelle diagrams with three different spacings 
obtained by SSA for stars No. 18 and 80. The top panels give the FT diagram. 
These panels are marked by the highest peak. The other panels are marked by the spacing 
used for getting the echelle diagrams. The highest peak of FT and the best solution of SSA do not agree.
} \label{fig13}
\end{figure}
 
A general comparison of FT spacing to our spacing values, both visual (VI) and algorithmic (SSA), 
reveals that the two methods (three approaches) do not yield a unique solution. There 
are cases when VI, SSA and FT spacings are the same (stars No. 2, 4, 5, 6, 11, 48 and 79) despite 
the spacings being around 2.2$\pm$0.1 d$^{-1}$ (stars No. 2 and 4) or 
around 1.7$\pm$0.2 d$^{-1}$  (stars No. 5 and 6) or around 3.5$\pm$0.1 d$^{-1}$ 
(stars No. 11, 48 and 79). As the echelle diagrams show, these stars have the simplest regular structure. 
There are cases when VI and SSA spacings are the 
same (stars No. 8, 12, 25 and 43), but the FT shows different spacings. There are cases, when
VI and FT spacings are the same (stars No. 19 and 24) or SSA and FT spacings are 
the same (stars No. 13, 23, 32 and 39). In Fig.~\ref{fig12} we present some 
characteristic examples of FT, representing the simplest cases (upper panels), 
the most complicated cases, when the decision which is the highest peak is hard (middle panels), 
and cases when FT shows a completely different spacing than VI and SSA (bottom panels). Our example for the visual inspection, star No 65, belongs to this group. 
We omitted the low-frequency region applying the Nyquist frequency to the FT.

We present the numbers of the spacings in 1 d$^{-1}$ bins for the 
different methods in Table~\ref{distrib}.   
The numbers in a bin are slightly different for VI and SSA, but FT shows a remarkably higher 
number in the 0-1 and 1-2 d$^{-1}$ region of the spacings.  
In a latter phase the 1-2 d$^{-1}$ bin was divided in to two parts to avoid the artifact 
of the lower limit of SSA for spacing (1.5 d$^{-1}$). VI and SSA have low numbers in 
the 1-1.5 d$^{-1}$ bin, while FT has much higher value.
The VI definitely interpreted such a spacing as a shift of the sequences. SSA has lower value probably 
due to the lower limit that we learned from the VI. In the 1.5-2.0 d$^{-1}$ bin the VI still has 
lower population but SSA and FT found a similar population. In both cases there is no additional 
search for the shifts of the sequences.

It is worthwhile to see how the different SSA spacings, when more are obtained, are related to the FT spacing.
We present two cases. The FT 
diagram and the echelle ridges where we used the different spacings are shown in Fig.~\ref{fig13}. 
Left panels belong to star No. 18, while right panels to star No. 80. The panels are labeled with 
the actual spacing value that we used for the calculation of the modulo values. 
The top panels show the FT. The second panels give the dominant SSA spacing resulting in the most straight 
echelle ridges, but the other values also fulfill the requirement of SSA. The FT agrees with one of 
the SSA spacings, but not necessarily with the dominant SSA spacing.

\begin{deluxetable}{rrrr}
\tablecaption{Spacing distributions \label{distrib}}  
\tablehead{
\colhead{Range} & \colhead{$N_{\mathrm {VI}}$} & \colhead{$N_{\mathrm {SSA}}$} & \colhead{$N_{\mathrm {FT}}$} }
\startdata
0-1 & $-$ & $-$ & 7 \\
1-2 & 5 & 16 & 25 \\
(1-1.5 & $-$ & 2 & 13) \\
(1.5-2 & 5 & 14 & 12) \\
2-3 & 35 & 31 & 23 \\
3-4 & 26 & 19 & 16 \\
4-5 & 3 & 6 & 9 \\
6-7 & 1 & 3 & 3 \\
7- & $-$ & 3 & $-$ 
\enddata                                               
\tablecomments{
Distribution of spacings obtained by different methods in 1 d$^{-1}$ bins.
The columns show the spacing range and the number of the spacings found by
the methods VI, SSA, and FT within the given range.
}
\end{deluxetable} 

We conclude that the different methods (with different requirements) are able to 
catch different regularities among the frequencies. The different spacing values 
are not a mistake of any method but the methods are sensitive to different regularities. 
The VI and SSA concentrate on the continuous sequence(s), while the FT is sensitive to 
the number of similar frequency differences, disregarding how many sequences are among the frequencies. 
When we have a second sequence with a 
midway shift, then the FT shows it, instead of the spacing of a single sequence. 
The spacing of a single sequence will be double the value of the highest peak in FT.

If the shifts of the sequences are asymmetric, the FT shows a low and a larger value with 
equal probability. When we have many peaks in the FT, then it reflects that we have many echelle ridges with 
different shifts with respect to each other. The sequence method helps to explain the fine structure of the FT.
 
\section{Test for refusing artifacts and confirmation of sequences}\label{test}

The comparison of spacing obtained by three different approaches results in a satisfactory 
agreement if we consider the different requirements. However, the spacing is the only point 
where we are able to compare them, since this is the only 
output of FT. We cannot compare the unexpectedly large number of echelle ridges (sequences), 
since we identified them for the first time. What we can do and what we really did, is to make 
any test that can rule out some possible artifacts and confirm the existence of so many sequences 
with almost equal spacing in $\delta$ Scuti stars.

(1) We started with a very basic test. Can we get the echelle ridges as a play of 
randomness on normally distributed frequencies? Three tests, one-dimensional Kolmogorov-Smirnov (K-S) test, 
Cram\'er-von~Mieses test, and the $\chi^2$-test 
were applied to our frequency list for the stars and to randomly generated frequency lists.
The frequency distribution of 14 stars showed significant differences from the 
normal distribution, but in the mathematical sense  most of our frequency list 
proved to be randomly distributed. 
The surprising mathematical test inspired more check.

The classical K-S test and its more sensitive refinements such as
Anderson-Darling
or Cram\'er-von~Mieses tests are successfully applied for small samples.
These tests are indeed the suggested tools for small element ($\sim 20$) samples.
Our frequency lists have 9-68 elements; the average value is 32.8. We prepared a
30-element equidistantly distributed artificial frequency list. In our phrasing all
the 30 frequencies build one single sequence. None of the tests, however, found
significant differences from the randomness. If we increase the number of
our synthetic data points and we reach 100-200 elements (depending on the used test)
the tests detect the structure, viz. the significant (95\%)
difference from the normal distribution.

As an additional control case, we tested 30 frequencies of a pulsating model of
FG Vir (discussed in paper Part I). All tests revealed that the model frequencies
($l$=0, 1, and 2) were also randomly distributed, although these frequencies were
a result of a pulsation code and a sequence of grouped frequencies was reported
for FG Vir \citep{Breger05}. Adding the rotational triplets and
multiplets (64 frequencies) to the
list (altogether 94 frequencies), the tests proved a significant
difference from the normal distribution.
We conclude that these statistical tests would give correct results
for our specific distributions
only if we had two to four times more data points than we have. The
present negative results have no
meaning; they are only small sample effects. In other words, such
global statistical tests
are not suitable tools for detecting or rejecting any structures in
our frequency lists.

(2) If the echelle ridges that we found were coincidences only, we could find
similar regularities for random frequency distribution as well.
Checking this hypothesis we have chosen three stars
which represent well our results:
the stars No. 39, 10 and 92 show a single sequence with 6 frequencies,
four sequences with 21 frequencies (the average length of
a sequence is 5.25), and 7 sequences with 35 frequencies
(average length = 5), respectively.
We prepared 100 artificial data sets for each
of these stars. The data sets contain random numbers as frequencies
within the interval of the real frequency intervals. The number
of the random ``frequencies" is the same as the number of the
real frequencies. The real star amplitudes are randomly assigned
to the synthetic frequencies.
We run the SSA on these synthetic data with the same parameters as
we used for real data. 
We found the following results.

We compared two parameters of the test and real data: the total number of
frequencies located on echelle ridges, and the average length of the sequences.
In the most complex case (star No.~92) we did not find a regular structure in the 
simulated data, for which the total number of frequencies located on the echelle ridges is
as high as in the real star (35).
In the two simpler cases  
only 5\% (for star No.~39) and 2\%  (for star No.~10) of the echelle ridges proved to be as long as in the real stars. 

These Monte Carlo tests show that the coincidence
as an origin of few of the echelle ridges that we found in our sample stars cannot be ruled out
completely,  but the probability of such a scenario is low ($<$5\%)
 and depends highly on the complexity of the echelle ridges (the more echelle ridges the lower the 
probability). This could concern, in the case of our sample, a
maximum of one to three stars.

(3) Obviously a basic test was whether any regularity can be caused by the instrumental effects 
(after removing most of them) and whether data sampling resulted in the systematic spacing of the frequencies?
The well-known effect from the ground-based observations (especially from single sites) 
is the 1, 2, $\dots$, d$^{-1}$ alias structure around the pulsation frequencies. 
In this case, we worked on continuous observation with the CoRoT space telescope. 
In principle, it excludes the problem of alias structure, but the continuity is interrupted 
from time to time by the non-equal long gaps caused by passing through the South Atlantic Anomaly (SAA). 
In the spectral window pattern the only noticeable alias peak is at 2.006 d$^{-1}$ and
sometimes an even lower peak around 4 d$^{-1}$. The expected alias structure around any pulsation peak 
is only 2 percent. A test on a synthetic light curve was presented 
by \citet{Benko15}. Comparison of the equally spaced and gapped data shows 
no difference in the frequencies. The requirement for a sequence
containing four members is, at least, a quintuplet structure of the alias peaks around the 
frequencies of the highest amplitude, which is very improbable for the CoRoT data. We may conclude 
that our sequences are not
caused by any alias structure of the CoRoT data. Table~\ref{bigtable} contains some spacings
with near integer value, but in most cases different methods yielded different
values. In an alias sequence we must have strictly equal spacing and mostly
only one echelle ridge.

(4) The linear combination of the higher amplitude modes creates a systematic arrangement 
of the frequencies reflecting the spacing between the highest amplitude modes.
A high amplitude $\delta$ Scuti star, CoRoT~101155310 \citep{Poretti11} was
used as a control case for two reasons. No systematic spacing was found for the 13 independent 
frequencies by our SSA algorithm which means the star does not show any instrumental effects 
discussed in the previous paragraph. To complete the list with the linear
combination, our algorithm found a dominant spacing around 2.67 d$^{-1}$ which is 
near the frequency difference of the highest amplitude modes.

Our visual inspection and algorithmic search were based on the 
investigation of the spacing of the peaks of the highest amplitude. It was 
necessary to check the frequency list for linear combinations. 
Half of our targets (44) showed linear combinations, with one (15) or two (12) 
$f_{\mathrm a}+f_{\mathrm b}=f_{\mathrm c}$
connections. In some cases (stars No. 21, 54, 66, 78, 7, 74 and 8) 9-14 linear
combination frequencies were found. Comparing these to the frequencies in the
echelle ridges, we found that the linear combinations were not included in the
echelle ridges. There is only a single case (star No. 71.) where the echelle
ridge at around 0.18 modulo value contains three members of a linear
combination.
In other cases, only two members fit the echelle diagrams. Star No. 38 is a critical case, 
where by omitting a member of the linear combination frequencies, we have to delete the single echelle ridge. 
We conclude that the echelle structure is not seriously modified in the other targets.

All echelle frequencies connected to linear combinations in our stars were compared. 
The frequencies are different from star to star, so the connection between the frequencies does not 
have any technical origin.

(5) To keep the human brain's well-known property in check, namely that it searches 
everywhere for structures (visual inspection) or any artifact in the algorithm, we used
well-known $\delta$ Scuti stars as test cases. Spacing of consecutive radial orders 
were published for different $\delta$ Scuti stars: 44 Tau \citep{Breger08}, 
BL Cam \citep{Rodriguez07}, FG Vir \citep{Breger05}, summarized by \citet{Breger09} 
and KIC 8054146 \citep{Breger12}. 
We checked these stars by our SSA algorithm to see whether we would find similar results or not.

\begin{deluxetable}{crrrr}
\tablecaption{Comparison of spacings \label{comp} }  
\tablehead{
\colhead{Star} & \colhead{SP$_{\mathrm p}$} 
& \colhead{SP$_{\mathrm A}$} &  \colhead{EF$_{\mathrm A}$} &  \colhead{SN$_{\mathrm A}$} \\ 
\colhead{}  &   \colhead{(d$^{-1}$)} & \colhead{(d$^{-1}$)} & \colhead{} & \colhead{} }
\startdata
44 Tau & 2.25 & 4.62 & 22 & 5 \\
BL Cam & 7.074 & 7.11 & 8 & 2 \\
FG Vir & 3.7 & 3.86 & 15 & 3 \\
KIC 8054146 & 2.763 & 2.82,3.45 & 7,12 & 1,2 
\enddata                                               
\tablecomments{The first two columns contain the star name and published spacing 
SP$_{\mathrm p}$. The last three columns show the results of our SSA search: spacing (SP$_{\mathrm A}$), 
the total number of
frequencies in all sequences (EF$_{\mathrm A}$), and the number of found sequences (SN$_{\mathrm A}$), respectively.}
\end{deluxetable} 

We summarize the results in Table~\ref{comp}. 
The published and the SSA spacings are in good agreement (2nd and 3rd columns). 
For 44 Tau we found double the value of the published spacing. In the case of KIC 8054146 we 
found a second spacing by SSA in addition to the first matching spacing. We also present the number 
of frequencies involved in the sequences and the number of sequences (4th and 5th columns).

We conclude that our algorithm finds the proper spacing of the data and the VI 
does not simply reflect the human brain. We confirmed that the echelle ridges belong to the pulsating stars 
and reflect the regularities connected to the stars.

\begin{figure}
\includegraphics[width=8cm]{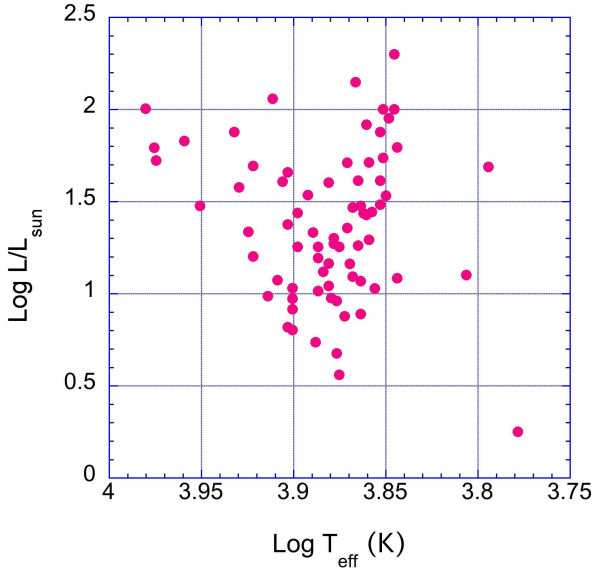}
\caption[]{
Theoretical HR diagram derived from the parameters obtained from the AAO spectroscopy. 
The location of the targets was used to derive the estimated rotational velocity and rotational frequencies
} \label{fig19}
\end{figure}

\section{Rotation-pulsation connection}

The basic problem of the mode identification in $\delta$ Scuti stars is partly the lack 
of regular arrangement of the frequencies predicted by the theory. Further complication is 
caused by the rotational splitting of the non-radial modes, especially for fast rotating stars.
Application of our sequence search method for $\delta$ Scuti stars revealed an unexpectedly 
large number of echelle ridges in many targets. Knowing the regular spacing of the frequencies 
we wonder whether we can find a connection between the echelle ridges and the rotational 
frequency of the stars.

\subsection{Estimated rotational velocities}

We do not have rotational velocity independently measured for our targets. 
Nowadays the space missions have enormously increased the number of stars investigated photometrically 
with extreme high precision, but the ground-based spectroscopy cannot keep up with this increase. 
However, for our targets we have at least AAO spectroscopy for classification 
purposes \citep{Guenther12, Sebastian12}.

Based on the AAO spectroscopy one of us \citep{Hareter13} derived the 
T$_{\mathrm{eff}}$ and $\log g$ values for our sample  used the same rotational 
velocity (100 km$^{-1}$) for all the stars (see Table~\ref{bigtable}). The error bars are
also given in \citet{Hareter13}. For giving insight to the error of AAO spectroscopy we present the
most typical range of errors for T$_{\mathrm{eff}}$ and $\log g$. For 70\% of the stars the 
error of T$_{\mathrm{eff}}$ falls in the range of 50-200 K. In a few cases ($\le 9$) the errors are 
over 1000 K. For  $\log g$  the typical range is 0.2-0.8, that represents of the 83\% of the stars. 
The physical parameters were used to plot our targets on the theoretical HR diagram as 
shown in Fig.~\ref{fig19}. To get a more sophisticated knowledge on the rotation of our targets we 
followed the process of \citet{Balona15} who determined 10 boxes on the theoretical HR diagram. 
Using the catalog of projected rotational velocities \citep{Glebocki00} they determined 
the distribution of $v\sin i$ for each box. The true distribution of equatorial velocities in 
the boxes was derived by a polynomial approximation \citep{Balona75}. 
Using the characteristic equatorial rotational velocities of the boxes we derived the 
estimated equatorial rotational velocity and the rotational frequency for each target 
presented in Table~\ref{est}. To obtain the rotational frequency, an estimate of the stellar radius is required; we
followed \citet{Balona15} in using the polynomial fit of \citet{Torres10} developed from studies of 
94 detached eclipsing binary systems plus $\alpha$ Cen.  The polynomial fit is a function 
of T$_{\mathrm{eff}}$, $\log g$, and [Fe/H]; we assume solar metallicity for our estimates.   
The radii calculated this way are also given in Table~\ref{est}. We also included the mass and 
the mean density in the table, calculating these using the \citet{Torres10} polynomial fit for mass 
and radius.

Although these are only estimated values, they allow us to compare the rotational 
frequency and the shifts between the sequences to search for a connection between them, if there is any.

\begin{deluxetable*}{rrrrrrrrrrrrr}
\tabletypesize{\scriptsize}
\tablecaption{Estimated sterllar properties \label{est} }  
\tablehead{
\colhead{Star} & \colhead{$R$} & \colhead{$V_{\mathrm{eq}}$} & \colhead{$\Omega_{\mathrm{rot}}$} & \colhead{$M$} & \colhead{$\rho$} & \colhead{}  &  
\colhead{Star} & \colhead{$R$} & \colhead{$V_{\mathrm{eq}}$} & \colhead{$\Omega_{\mathrm{rot}}$} & \colhead{$M$} & \colhead{$\rho$} \\ 
\colhead{}     & \colhead{(R$_\sun$)}            & \colhead{(km~s$^{-1}$)}         & \colhead{(d$^{-1}$)} &    \colhead{(M$_\sun$)}   & \colhead{(g~cm$^{-3}$)}                 & \colhead{}  &    
\colhead{}     & \colhead{(R$_\sun$)}            & \colhead{(km~s$^{-1}$)}         & \colhead{(d$^{-1}$)} &    \colhead{(M$_\sun$)}   & \colhead{(g~cm$^{-3}$)}
}
\startdata
1=55  &  3.911  &  80  &  0.404  & 2.046 & 0.0482 & &  49  &  1.936  &  130  &  1.327 & 1.725 & 0.3349  \\
2=66  &  3.985  &  110  &  0.545  & 2.528 & 0.0563 & &  50  &  2.176  &  110  &  0.999 & 1.728 & 0.2360  \\
3  &  9.641  &  80  &  0.164  & 3.044  &0.0048 &  &  51  &  3.414  &  80  &  0.463 & 1.976 & 0.0699  \\
4  &  2.340  &  70  &  0.591  & 1.776  & 0.1952 &  &  52  &  1.745  &  130  &  1.472 & 1.764 & 0.4681  \\
5  &  2.409  &  70  &  0.574  &  1.677 & 0.1690 & &  53  &  5.402  &  110  &  0.402 & 2.726 & 0.0244  \\
6  &  3.335  &  140  &  0.829  & 2.527 & 0.0959 & &  54  &  5.797  &  100  &  0.341  & 2.486 & 0.0180  \\
7  &  6.370  &  80  &  0.248  & 2.519 & 0.0137 &  &  55=1  &  3.911  &  80  &  0.404 & 2.046 & 0.0482  \\
8=92  &  3.540  &  110  &  0.614  & 2.248 & 0.0714 &  &  56  &  3.347  &  110  &  0.649  & 2.013  & 0.0756  \\
9  &  5.726  &  100  &  0.345  & 2.428 & 0.0182 &  &  62  &  2.311  &  150  &  1.282  & 2.184  & 0.2492 \\
10  &  2.679  &  110  &  0.811  & 1.840 & 0.1348 &  & 63  &  3.448  &  110  &  0.630  & 2.014 & 0.0692 \\
11=81  &  1.347  &  130  &  1.907  & 1.660 & 0.9574 &  &  65  &  4.008  &  100  &  0.493 & 2.146 & 0.0470  \\
12  &  1.928  &  150  &  1.537  &  1.920 & 0.3770 & &  66=2  &  3.985  &  110  &  0.545 & 2.528 & 0.0563  \\
13=74  &  6.651  &  100  &  0.297  & 2.588 & 0.0124 &  &  67  &  1.767  &  130  &  1.454  & 1.809 & 0.4620 \\
14=96  &  2.222  &  130  &  1.156  & 1.800 & 0.2309 &  &  68  &  3.304  &  140  &  0.837 & 2.204 & 0.0860  \\
15  &  1.352  &  130  &  1.899  &  1.674 & 0.9532 & &  69  &  1.297  &  90  &  1.371  & 1.541 & 0.9955 \\
18  &  1.143  &  90  &  1.555  & 1.499 & 1.4126 &  &  70  &  1.633  &  130  &  1.573  & 1.734 & 0.5611 \\
19  &  1.796  &  90  &  0.990  &  1.669 & 0.4055 & &  71  &  3.702  &  140  &  0.747  & 2.767 & 0.0768 \\
20  &  2.986  &  110  &  0.728  & 3.073  & 0.1626 &  & 72  &  7.355  &  100  &  0.269  & 2.812  & 0.0100 \\
21  &  1.825  &  130  &  1.407 & 1.721 & 0.3988 &  &  73  &  2.406  &  130  &  1.068  & 1.875  & 0.1897 \\
22  &  1.248  &  40  &  0.633  & 1.153 & 0.8345 &  &  74=13  &  6.651  &  100  &  0.297  & 2.588 & 0.0124 \\
23  &  6.039  &  80  &  0.262  & 2.151 & 0.0138 &  &  75  &  2.913  &  40  &  0.271  & 1.629 & 0.0929 \\
24  &  2.282  &  110  &  0.952  & 1.897 & 0.2247 & &  76  &  2.907  &  110  &  0.748  & 1.923  & 0.1103 \\
25  &  2.220  &  150  &  1.335  & 2.015 & 0.2595 & &  77  &  4.235  &  80  &  0.373  & 2.131 & 0.0395 \\
26  &  2.547  &  150  &  1.163  & 1.870 & 0.1593 & &  78  &  4.910  &  70  &  0.282  & 2.260 & 0.0269 \\
27  &  1.668  &  130  &  1.540  & 1.616 & 0.4903 &  &  79  &  2.161  &  70  &  0.640  & 1.704 & 0.2378 \\
28  &  5.430  &  80  &  0.291  & 2.318 & 0.0204 & &  80  &  3.347  &  110  &  0.649  & 2.013 & 0.0756 \\
29  &  2.096  &  110  &  1.037  & 3.232 & 0.4947 &  &  81=11  &  1.347  &  130  &  1.907 & 1.660 & 0.9574 \\
30  &  3.649  &  80  &  0.433  & 2.005 & 0.0581 & &  82  &  1.320  &  130  &  1.945  & 1.596 & 0.9770 \\
31  &  2.135  &  70  &  0.648  & 1.664 & 0.2409 &  &  83  &  2.558  &  140  &  1.081  & 1.997 & 0.1680 \\
32  &  2.851  &  70  &  0.485  & 1.853 & 0.1126 &  &  84  &  1.759  &  90  &  1.011  & 1.601 & 0.4146 \\
33  &  6.839  &  100  &  0.289  & 2.583 & 0.0114 & &  86  &  3.307  &  110  &  0.657  & 1.967 & 0.0766 \\
34  &  2.962  &  140  &  0.934  & 2.524 & 0.1368 & &  87  &  4.360  &  100  &  0.453  & 2.255 & 0.0383 \\
35  &  2.536  &  110  &  0.857  & 1.853  & 0.1601 &  &  88  &  4.610  &  100  &  0.429   & 2.242  & 0.0322 \\
36  &  1.530  &  130  &  1.679  & 1.707 & 0.6716 &  &  89  &  3.679  &  100  &  0.537  & 2.158  & 0.0611 \\
37  &  3.315  &  110  &  0.656  & 1.976  & 0.0764 &   &  90  &  2.083  &  130  &  1.233  & 1.777 & 0.2770 \\
38  &  2.640  &  110  &  0.823  & 1.893 & 0.1449 & &  91  &  3.234  &  110  &  0.672  & 2.113 & 0.0880 \\
39  &  1.530  &  130  &  1.679  & 1.707 & 0.6716 &  &  92=8  &  3.540  &  110  &  0.614  & 2.248 & 0.0714 \\
40  &  2.823  &  110  &  0.770  &  2.038 & 0.1276 & &  93  &  1.805  &  130  &  1.423  & 1.684 & 0.4035 \\
43  &  2.754  &  140  &  1.004  & 2.456  & 0.1655 & &  94  &  2.243  &  110  &  0.969  & 1.832 & 0.2288 \\
45  &  1.562  &  150  &  1.897  & 1.779  & 0.6573 & &  95  &  2.594  &  110  &  0.838  & 1.937 & 0.1564 \\
47  &  2.861  &  140  &  0.967  & 2.220  & 0.1335 &   &  96  &  2.222  &  130  &  1.156  & 1.800  & 0.2309 \\
48  &  3.387  &  140  &  0.817  & 2.315 & 0.0839 &  & &  &    &    &    &    
\enddata                                               
\tablecomments{The table contains the running number, the radius of the star,
        the estimated rotational velocity, estimated rotational frequency, mass, and mean density. The radius and mass used in these estimates were calculated from the spectroscopic parameters using the formulas of \citet{Torres10}, assuming solar metallicity.  The same parameters for stars after running numbers 48 can be found in the 7th, 8th, 9th, 10th, 11th, and 12th columns.
}
\end{deluxetable*} 

\subsection{Echelle ridges and rotation}

We have three parameters that we can compare for our targets, namely the shift of the sequences, 
the rotational frequencies derived, and the spacing, or in some cases the spacings.

\subsubsection{Midway shift of the sequences}

\begin{deluxetable}{llll}
\tablecaption{Midway shifts \label{midway} }  
\tablehead{
\colhead{Star} & \colhead{Spacing}  & \colhead{No. of ridges} &  \colhead{Mod. of ridges}  \\ 
\colhead{}     & \colhead{(d$^{-1}$)} & \colhead{(\%)} & \colhead{}  
} 
\startdata
7    & 1.795 & 1-2 {\it (8)} & 0.36-0.89 \\
8    & 2.481 & 7-8 {\it (1)} & 0.07-0.58 \\
     &       & 5-4 {\it (1)} & 0.29-0.79 \\
10   & 2.614 & 4-3 {\it (3)} & 0.24-0.76 \\
13   & 2.674 & 2-4 {\it (5)} & 0.90-0.37 \\
18   & 6.001 & 1-2 {\it (9)}  & 0.90-0.35 \\
20   & 1.478 & 1-2 {\it (1)} & 0.06-0.57 \\
27   & 5.995 & 2-3 {\it (2)} & 0.11-0.62 \\
28   & 2.655 & 3-5 {\it (6)} & 0.93-0.22 \\
     &       & 4-2 {\it (11)} & 0.05-0.60 \\
     &       & 3-6 {\it (12)} & 0.93-0.37 \\
29   & 2.389 & 6-4 {\it (0)} & 0.77-0.26  \\
32   & 1.671 & 1-3 {\it (6)} & 0.63-0.17 \\
33   & 2.396 & 4-3 {\it (1)} & 0.67-0.19 \\
54   & 2.300 & 1-2 {\it (3)} & 0.54-0.03 \\
66   & 2.406 & 4-2 {\it (0)} & 0.93-0.43 \\
71   & 3.495 & 1-4 {\it (3)} & 0.65-0.14 \\
     &       & 3-2 {\it (1)} & 0.85-0.34 \\
72   & 2.249 & 1-3 {\it (4)} & 0.16-0.68 \\
     &       & 2-4 {\it (9)} & 0.43-0.89 \\
74   & 2.940 & 7-5 {\it (5)} & 0.64-0.15 \\
     &       & 8-2 {\it (2)} & 0.77-0.28 \\
76   & 1.772 & 1-3 {\it (9)} & 0.79-0.25 \\
     &       & 2-3 {\it (5)} & 0.73-0.25 \\
77   & 2.521 & 1-2 {\it (3)} & 0.51-0.04 \\
78   & 2.392 & 6-4 {\it (5)} & 0.75-0.22 \\
87   & 1.867 & 3-2 {\it (5)} & 0.05-0.54 \\
92   & 2.576 & 1-2 {\it (7)} & 0.85-0.31 \\
     &       & 3-5 {\it (1)} & 0.13-0.62 \\
     &       & 7-6 {\it (4)} & 0.19-0.70 \\
96   & 2.464 & 1-7 {\it (5)} & 0.49-0.02 \\
     &       & 6-5 {\it (3)} & 0.76-0.27 \\
     &       & 3-2 {\it (3)} & 0.92-0.41 
\enddata                                               
\tablecomments{
 The table contains the running numbers,
the spacing, the numbering of the echelle ridges and the modulo value of the echelle ridges
for identification in Figs.~\ref{fig3}-\ref{fig10}. The ratio of the shift of the sequences and half of the spacing is given by italics in 3rd column.
}
\end{deluxetable}

In the framework of the sequence search method we derived the shifts between each pair of 
sequences as we described in Sec.~\ref{visual}. The independent shifts were averaged for the members in 
the sequence. In the rest of the paper we refer to the average value when we mention the shift. 
There are two expectations for the shifts. Similar to the spacing in the asymptotic regime, 
the sequences of the consecutive radial orders of the different $l$ values are shifted relative 
to each other. For example the $l$=0 and $l$=1 radial orders are shifted to midway between the 
large separation in the asymptotic regime. The other possible expectation for the shift is the 
rotational splitting. We checked the shifts of each target for both effects.

Of course, we have shifts only when we found more that one echelle ridge. 
Only one echelle ridge was found in 20 stars. In 34 stars we have no positive result for 
the midway shift. However, we found shifts with half of the regular spacing (shifted to midway) 
in 22 stars. We present them in Table~\ref{midway}. The table contains the running numbers, 
the spacing, the numbering of the echelle ridges and the modulo value of the echelle ridges 
for identification in Figs.~\ref{fig3}-\ref{fig10}. To follow how precise the midway shift is, we give the deviation 
in percentage by italics. In some stars there are two pairs with a midway shift 
(stars No. 8, 71, 72, 74 and 76), while in stars No. 28, 92 and 96 three pairs appear 
with a midway shift compared to the spacing. Of course, it could happen that the shift to 
midway represents a 1:2 ratio of the estimated rotational frequency and the spacing, but 
we mentioned them independently as a similarity to the behavior in the asymptotic regime. 
In general the ratio of the dominant spacing to the rotational frequency is in the 1.5-4.5 interval 
for most of our targets (52 stars).

\subsubsection{Shift of sequences with the rotational frequency}

The pulsation-rotation connection appears in a prominent way when one, two or even more 
shifts between pairs of the echelle ridges agree with the rotational frequency. We found 31 stars 
where a doublet, triplet or multiplet appears with a splitting near the rotational frequency. 
In Table~\ref{doublet} we give the running number of stars, the estimated rotational frequency, 
the shifts between the sequences, the numbering of echelle ridges connected to each other, 
and the modulo values of these echelle ridges for identification on Figs~\ref{fig3}-\ref{fig10}.

\begin{deluxetable*}{lllll}
\tablecaption{Doublets, triplets and multiplets \label{doublet} }  
\tablehead{
\colhead{Star} & \colhead{$\Omega_{\mathrm{rot}}$} & \colhead{Shift}  & \colhead{No. of ridges} &  \colhead{Mod. of ridges}  \\ 
\colhead{}     & \colhead{(d$^{-1}$)}               & \colhead{(d$^{-1}$)} & \colhead{}            & \colhead{}  
} 
\startdata
1 & 0.404 & 0.455 {\it (13)} & 4-5 & 0.17-0.39 \\
     &       & 0.482 {\it (19)} & 3-2 & 0.52-0.73 \\
7    & 0.248 & 0.224 {\it (11)} & 6-2 & 0.78-0.89 \\
     &       & 0.224-0.563 {\it(11-14*)} & 6-2-5 & 0.78-0.89-0.22 \\
8    & 0.614 & 0.521-0.551 {\it (18-11)} & 6-8-4 & 0.37-0.58-0.79 \\
     &       & 0.325-0.355 {\it (6*-16*)} & 5-2-8 & 0.29-0.43-0.58 \\
9    & 0.345 & 0.664 {\it (4*)} & 3-2 & 0.91-0.14 \\
10   & 0.811 & 0.874 {\it (8)} & 1-2 & 0.16-0.50 \\
     &       & 0.874-0.662 {\it (8-23)} & 1-2-3 & 0.16-0.50-0.75 \\
13  & 0.297 & 0.317 {\it (7)} & 1-2 & 0.78-0.90  \\
18  & 1.555 & 1.430 {\it (9)} & 1-4 & 0.90-0.15 \\
    &       & 1.649 {\it (6)} & 3-2 & 0.09-0.35 \\
    &       & 1.315 {\it (18)} & 4-2 & 0.15-0.35 \\
    &       & 1.430-1.315 {\it (9-18)} & 1-4-2 & 0.90-0.15-0.35 \\
20 & 0.727 & 0.886 {\it (22)} & 3-4 & 0.30-0.75 \\
27 & 1.540 & 1.533 {\it (0)} & 1-4 & 0.90-0.16 \\
28 & 0.291 & 0.308 {\it (6)} & 3-4 & 0.93-0.05 \\
29 & 1.037 & 0.997-1.041 {\it (4-0)} & 1-6-2 & 0.35-0.77-0.20 \\
   &       & 0.546-0.451 {\it (5*-15*)} & 1-7-6 & 0.35-0.56-0.77 \\
30 & 0.433 & 0.459 {\it (6)} & 2-1 & 0.77-0.91 \\
   &       & 0.901 {\it (4*)} & 4-2 & 0.51-0.77 \\
31 & 0.648 & 0.587 {\it (10)} & 1-3 & 0.64-0.86 \\
32 & 0.485 & 0.478 {\it (1)} & 1-4 & 0.63-0.77 \\
   &       & 0.571 {\it (18)} & 3-2 & 0.17-0.33 \\
33 & 0.289 & 0.251 {\it (15)} & 5-3 &  0.08-0.19  \\
49 & 1.327 & 1.568 {\it (18)} & 1-3 &  0.39-0.05  \\
53 & 0.402 & 0.428 {\it (6)} & 4-1 & 0.48-0.58  \\ 
   &       & 0.361 {\it (11)} & 2-3 & 0.68-0.80 \\ 
54 & 0.341 & 0.611 {\it (12*)} & 1-3 & 0.54-0.79  \\ 
55 & 0.404 & 0.818 {\it (1*)} & 2-5 & 0.45-0.78  \\ 
   &       & 0.729-0.818 {\it (11*-1*)} & 3-2-5 & 0.17-0.45-0.78  \\ 
62 & 1.282 & 1.277 {\it (0)} & 1-2 & 0.61-0.92 \\
63 & 0.630 & 0.611 {\it (3)} & 4-2 & 0.67-0.87 \\ 
66 & 0.545 & 0.563 {\it (3)} & 1-4 & 0.69-0.93 \\
   &     & 0.269-0.301 {\it (1*-10*)} & 1-3-4 & 0.69-0.80-0.93 \\
71 & 0.747 & 0.654 {\it (16)} & 1-3 & 0.65-0.85 \\
   &       & 0.672 {\it (11)} & 4-2 & 0.14-0.34 \\
   &       & 0.859 {\it (15)} & 5-6 & 0.75-0.02 \\
72 & 0.269 & 0.603-0.563 {\it (12*-5*)} & 1-2-3 & 0.16-0.43-0.68 \\
73 & 1.068 & 1.247 {\it (17)} & 2-1 & 0.20-0.57  \\
74 & 0.297 & 0.359-0.356-0.346 {\it (21-20-17)} & 4-3-5-2 & 0.91-0.05-0.15-0.28 \\
76 & 0.748 & 0.811-0.844 {\it (8-13)} & 1-3-2 & 0.79-0.25-0.72 \\
78 & 0.282 & 0.364-0.345-0.297-0.361 {\it (29-22-5-28)} & 2-5-4-1-3 & 0.93-0.11-0.22-0.35-0.50 \\
84 & 1.011 & 0.506 {\it (0*)} & 4-2 & 0.42-0.60 \\
86 & 0.657 & 0.610 {\it (8)} & 3-1 & 0.26-0.32 \\
87 & 0.453 & 0.525-0.933-0.406 {\it (16-3*-12)} & 4-2-1-3 & 0.17-0.54-0.78-0.05 \\
88 & 0.429 & 0.357-0.322 {\it (20-33)} & 4-2-1 & 0.28-0.42-0.54 \\
92 & 0.614 & 0.567 {\it (8)} & 4-7 & 0.96-0.19 \\
   &       & 0.605-0.699 {\it (1-14)} & 5-1-3 & 0.62-0.85-0.13 \\
   &       & 0.766-0.605-0.699 {\it (25-1-14)} & 2-5-1-3 & 0.31-0.62-0.85-0.13 \\
93 & 1.423 & 1.285 {\it (11)} & 4-1 & 0.78-0.10 \\
   &       & 1.159-1.285 {\it (23-11)} & 3-4-1 & 0.51-0.78-0.10 \\
96 & 1.156 & 1.123-1.254 {\it (3-8)} & 1-3-2 & 0.49-0.92-0.41 
\enddata                                               
\tablecomments{The table contains the running number, the estimated rotational velocity,
 the shifts between the rotationally connected echelle ridges, 
the numbering of echelle ridges connected rotationally, and the modulo value of the echelle ridges 
for identification purpose on Figs.~\ref{fig3}-\ref{fig10}.
}
\end{deluxetable*} 

Of course, we may not expect that the estimated rotational frequency and the split (shift) 
of the doublet and the triplet components agree to high precision. As a guideline we 
followed \citet{Goupil00} who derived about 30\% deviation in the split of the component from 
the equally-spaced splitting. We accepted the doublets, triplets and multiplets if the deviation of 
the shifts is less than 20\% compared to the estimated rotational frequency. 
To follow how reliable are the doublets, triplets and multiplets we included the ratio of 
the actual shift and the estimated rotational frequency. In most cases presented in 
Table~\ref{doublet} the ratios are even less that 10\% (13 stars). We included some examples 
with higher than 20\%  representing triplets (stars No. 10, 8 and 93) or complete or 
incomplete multiplets (stars No. 78 and 92). 

For getting a complete view of the connection between the shifts and the estimated rotational 
frequencies, we included cases where shifts are twice (stars No. 9, 30, 54, 55 and 72) or 
half (stars No. 8, 29, 66 and 84) the value of the estimated rotational frequency. The deviations 
are marked by an asterisk in these cases. A missing component in an incomplete multiplet 
(star No. 87) is also marked by an asterisk.

The attached file to this paper with the flags allows the interested readers to 
derive the shifts between the pairs of the echelle ridges. The numbering of the flags agrees with 
the numbering in electronic table.

\subsubsection{Difference of spacings and the rotational frequency}  

\begin{deluxetable*}{rrrrrrrrr}
\tablecaption{Possible large separation \label{lsep}}
\tablehead{
\colhead{No} & \colhead{$SP_1$} & \colhead{$SP_2$} & \colhead{$SP_1-SP_2$} 
& \colhead{$\Omega_{\mathrm{rot}}$} & \colhead{(2)} & \colhead{(3)} & \colhead{(4)} & \colhead{(5)} \\
\colhead{} & \colhead{(d$^{-1}$)} & \colhead{(d$^{-1}$)} & \colhead{(d$^{-1}$)} 
& \colhead{(d$^{-1}$)} & \colhead{(d$^{-1}$)} & \colhead{(d$^{-1}$)} & \colhead{(d$^{-1}$)} 
& \colhead{(d$^{-1}$)}
}
\startdata
35 & 3.492 & 2.609 & 0.883 & 0.857 & *3.492 & 2.609 & 1.752 & 4.349 \\
45 & 3.306 & 1.407 & 1.889 & 1.897 & 3.306 & 1.407 & -- & *5.203 \\
47 (VI) & 2.525 & 1.597 & 0.928 & 0.967 & 2.525 & 1.597 & 0.63 & *3.492 \\
72 & 2.249 & 1.977 & 0.272 & 0.269 & 2.249 & 1.977 & *1.708 & 2.518 \\
73 & 3.416 & 2.417 & 0.999 & 1.068 & *3.416 & 2.417 & 1.349 & 4.484 \\
95 & 3.294 & 2.262 & 0.832 & 0.838 & *3.294 & 2.462 & 1.624 & 4.132 \\
\tableline
1 & 2.092 & 1.510 & 0.582 & 0.404 & *2.092 & 1.510 & 1.106 & 2.496 \\
22 & 2.598 & 1.877 & 0.721 & 0.633 & 2.598 & 1.877 & 1.244 & *3.231 \\
92 & 2.576 & 1.880 & 0.696 & 0.614 & *2.576 & 1.880 & 1.266 & 3.190 \\
96 (VI) & 3.387 & 2.429 & 0.958 & 1.156 & *3.387 & 2.429 & 1.273 & 4.543 \\
\tableline
9 & 3.506 & 2.784 & 0.722 & 0.345 & 3.506 & 2.784 & *2.439 & 3.851 \\
54 & 3.275 & 2.300 & 0.975 & 0.341 & 3.275 & 2.300 & *1.959 & 3.616
\enddata
\tablecomments{The columns contain the running numbers (No), 
the spacings, the difference of the spacings, the rotational frequency and the 
possible large separations in agreement with the Equations (2), (3), (4), and (5).}
\end{deluxetable*}

There are 25 stars in our sample where SSA found more than one spacing between 
the frequencies (see Table~\ref{bigtable}). 
Based on the results obtained for the model frequencies of FG Vir, namely that one 
of the spacing agrees with the large separation and the other one with the sum of the 
large separation and the rotational frequency, we generalized how to get the large separation if 
none of the spacing represent the large separation itself but both spacings are the combination of 
the large separation and the rotational frequency (Part I paper).
We recall the equations:
\begin{eqnarray}
SP_1 & =& \Delta\nu, \ {\mathrm {and}} \ SP_2 = \Delta\nu - \Omega_{\mathrm{rot}}, \\
SP_2 & =& \Delta\nu,  \ {\mathrm {and}} \ SP_1 = \Delta\nu + \Omega_{\mathrm{rot}}, \\
SP_1 & =& \Delta\nu + 2\cdot\Omega_{\mathrm{rot}}, \ {\mathrm {and}} \ SP_2 = \Delta\nu + \Omega_{\mathrm{rot}}, \\
SP_2 & =& \Delta\nu - 2\cdot\Omega_{\mathrm{rot}}, \ {\mathrm {and}} \ SP_1 = \Delta\nu - \Omega_{\mathrm{rot}},
\end{eqnarray}
where, $SP_1$ and $SP_2$ are the larger and smaller values of the spacings, respectively,
found by SSA, $\Delta\nu$ is the large separation in the traditionally used sense,
and $\Omega_{\mathrm{rot}}$ is the estimated rotational frequency.

The four possible value of the large separation ($\Delta\nu$) are (2) $\Delta\nu = SP_1$, (3) 
$\Delta\nu = SP_2$ (4) $\Delta\nu =  SP_2 - \Omega_{\mathrm {rot}}$ or 
(5) $\Delta\nu =  SP_1 + \Omega_{\mathrm {rot}}$.  
We applied these equations to the $SP_1$ and $SP_2$ spacings of CoRoT 102675756, the star No. 72 of our sample in Part I paper. 
Obtaining the possible values of the large separation, we plotted them on the mean density versus large 
separation diagram, along with the relation derived using stellar models by \citet{Suarez14}.  
We concluded that the most probable value of the large separation is the closest one to the relation.

We applied this concept in this paper to our targets in which the difference of the spacings agrees 
with the estimated rotational frequency exactly, or nearly, or in which the spacing difference is twice 
or three times of the rotational frequency. We mentioned the latest group for curiosity, 
where special relation appears between the estimated rotational velocity. We emphasize that our results
are not forced to fulfill the theoretical expectation. To keep the homogeneity we everywhere used 
the $\Omega_{\mathrm{rot}}$, the estimated rotation frequency, to calculate the large separation 
not the actual difference of the spacings if we have any.
In addition to the SSA solutions, we included solution for 
two stars (No. 47 and 96) from the VI that agreed with the aforementioned requirements. 
We calculated the four possible large separations for these stars that we present in Table~\ref{lsep}. 
The three groups, concerning the agreement of the difference of the spacings and the rotational frequency, are divided by a line. 
The columns give the running number, $SP_1$, $SP_2$, $SP_1-SP_2$, $\Omega_{\mathrm{rot}}$ 
and the four possible large separations in agreement with the Equations (2), (3), (4) and (5). 
Fig.~\ref{figexact}.
shows the location of the best fitting large separations (marked by asterisk in Table~\ref{lsep}) on the mean 
density versus large separation diagram, along with the relation given by \citet{Suarez14}.  
The three groups are shown by different symbols, and the large separations obtained from different 
equations are marked by different colors. The stars with $\Delta\nu = SP_1$ (black symbols) 
perfectly agree with the middle part of the theoretically derived line. These are the stars with 
intermediate rotational frequency. The stars with higher and lower rotational frequency marked by 
blue and green symbols and derived by $\Delta\nu =SP_2 - \Omega_{\mathrm{rot}}$ 
and $\Delta\nu = SP_1 + \Omega_{\mathrm {rot}}$, respectively, deviate more from the theoretical line. 
The small black open circles represent $\Delta\nu = SP_2 - 2\cdot \Omega_{\mathrm {rot}}$ 
(next the green symbols) or $\Delta\nu = SP_1 + 2\cdot \Omega_{\mathrm {rot}}$ values 
(next the blue symbols).

\begin{figure}
\includegraphics[width=9cm]{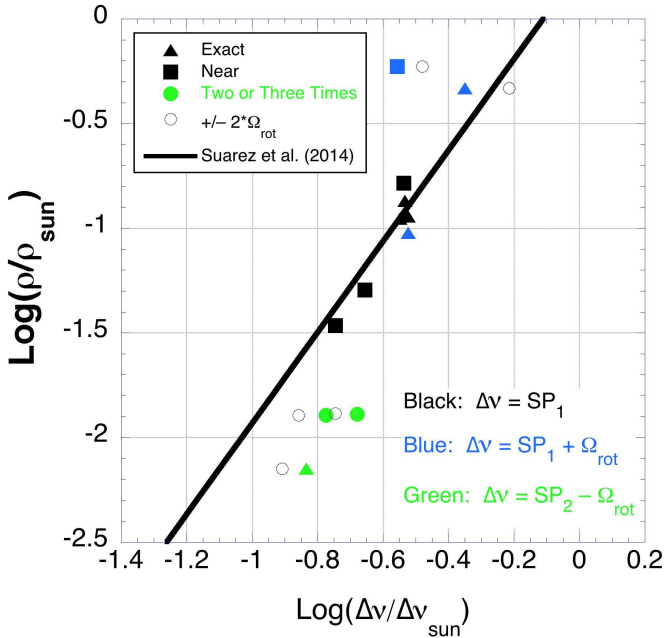}
\caption[]{
Location of the stars on the log mean density vs. log large separation diagram, along with 
the relation based on stellar models from \citet{Suarez14}. Three groups are those in which 
the difference of the spacings is: equal to the rotational frequency (triangle); near 
to that value (square); or twice or three times of the rotational frequency (circles) presented for
curiosity. The color code corresponds 
how the $\Delta\nu$ was calculated: black Eq.~(2), green Eq.~(4), and blue Eq.~(5). 
Open circles shows $\Delta\nu$ calculated with $\pm 2\cdot\Omega_{\mathrm{rot}}$.
\label{figexact}
}
\end{figure}

\begin{figure}
\includegraphics[width=9cm]{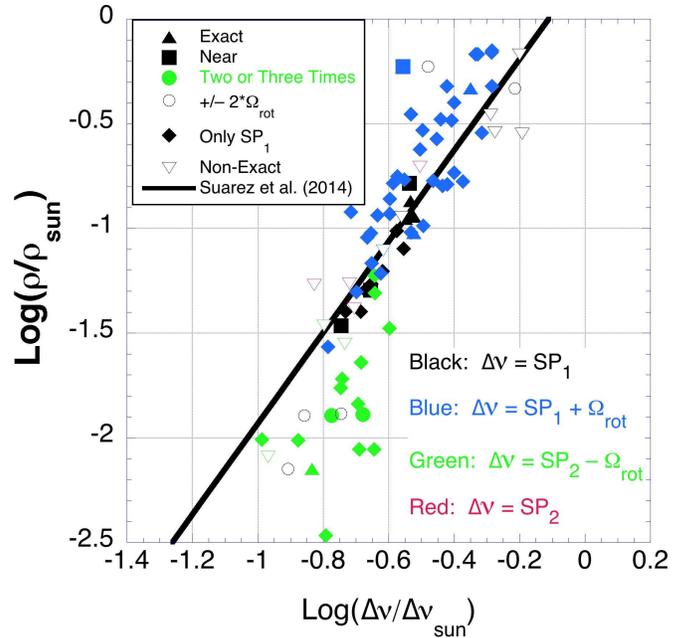}
\caption[]{
Location of the whole sample on the log mean density vs. log large separation diagram, 
along with the relation based on stellar models from \citet{Suarez14}. The new symbols represent 
the stars for which there is no agreement between the rotational frequency and the difference of 
the spacings (inverted triangle) or the stars with only one spacing (diamonds). 
The color code is the same as in the previous figure, with the addition of the red color 
corresponding to $\Delta\nu = SP_2$.
} \label{figall}
\end{figure}

We found numerical agreement between the difference of the spacings and the rotational frequency 
only in half of the stars (12 stars) for which SSA found more then one spacings. 
We do not know why we do not have numerical agreement for the other stars. 
A reason may be the uncertainties in estimated rotational velocity. 

Nevertheless, we proceeded to apply the conclusion based on Equations (2)-(5) deriving 
the possible large separation to the stars where we do not have an agreement (14 stars) and to 
the stars (53) where SSA found only one spacing.
Plotting in Fig.~\ref{figall} the best-fitting value of the large separation for both groups 
on the mean density versus large separation diagram along with the relation of \citet{Suarez14}, 
we found that the large separations are closely distributed along the \citet{Suarez14} line.
The figure contains not only the two new groups but the whole sample. 
Different symbols are used for the two groups (inverted triangle, and diamond, respectively) but 
the color code according to the calculation of $\Delta\nu$ is kept in the same sense 
as in Fig.~\ref{figexact}. The distribution of the whole sample is consistent. The stars 
with $\Delta\nu=SP_1$ agree with the middle part of the line, whether they fulfill the 
equations or not, although some stars appear with $\Delta\nu=SP_1$ on the upper part of the 
plot from group with two spacing. The deviation of the stars with higher and lower rotational 
frequency can be also noticed. We may have a slight selection effect in the lower $\Delta\nu$  
region due to the limitation of the spacing search at 1.5 d$^{-1}$.

We may conclude that we found an unexpectedly clear connection between the pulsational frequency spacings 
and the estimated rotational frequency in many targets of our sample. The tight connection 
confirms that our echelle ridges are not frequencies accidentally located along the echelle ridges. 
They represent the pulsation and rotation of our targets. 
Of course the well-determined rotational frequency for as large sample as we have would be 
needed to confirm the results with higher precision than we have here. However, this way of 
investigation seems to be a meaningful approach to disentangle the pulsation and rotation 
in the mostly fast rotating $\delta$ Scuti stars.

The frequencies along the ridges could be identified with the island modes in the ray 
dynamic approach, while frequencies widely distributed in the echelle diagrams could be the
chaotic modes. Both of them have observable amplitude in fast rotating stars, but only 
the island modes show regularity as the echelle ridges \citep{Ouazzani15}. For the authors 
it is not trivial to give a deeper interpretation of the results in the ray dynamic approach, 
but hopefully colleagues will interpret it in forthcoming papers.

\section{Summary}

We aimed to survey the possible regularities in $\delta$ Scuti stars on a large sample in order to determine whether or not 
we can use the regular arrangement of high precision space-based frequencies for mode identification.
Ninety stars observed by the CoRoT space telescope were investigated for regular spacing(s). We introduced 
the sequence search method with two approaches, the visual inspection and the algorithmic search. The visual 
inspection supported the parameter range and the tolerance value for quasi-equal spacing. The method proved 
to be successful in determining the dominant spacing and in finding sequences/echelle ridges in 77 stars stars 
from one up to nine ridges.
Compared to the spacings obtained by SSA and FT we concluded that the different methods 
(with different requirements) are able to catch different regularities among the frequencies. Not only does the 
spacing in a sequence represent regularity among the frequencies, but the shift of the sequences, too, can be found.

The sequence search method resulted in very useful parameters beside the most probable spacing, 
namely the shift of the sequences and the difference of the spacings.

The determination of the averaged shift between the pairs of echelle ridges opens a new field of investigation. 
With the comparison of the shift to the spacing, we determined one midway shift of at least one pair of the 
echelle ridges in 22 stars. Comparing the shifts to the estimated rotational frequency we recognized 
rotationally split doublets (in 21 stars), triplets (in 9 stars) and multiplets (in 4 stars) not only for a 
few frequencies, but for whole echelle ridges in $\delta$ Scuti stars that are pulsating in the non-asymptotic regime.

The numerical agreement between the difference of the spacings and the rotational frequency obtained 
for FG Vir (Part I paper) and in many of our sample stars (12) revealed a possibility for deriving the 
large separation ($\Delta\nu$) in $\delta$ Scuti stars pulsating in the non-asymptotic regime.
Generalized to those stars for which there is no numerical agreement between the difference 
of the spacings with the rotational frequency (14), or for which only one spacing was obtained 
by SSA (53), we found an arrangement of each target along the theoretically determined mean density 
versus large separation diagram \citep{Suarez14} calculating the $\Delta\nu$ as $\Delta\nu = SP_1$, 
$\Delta\nu = SP_2$, $\Delta\nu = SP_2 -\Omega_{\mathrm{rot}}$ and $\Delta\nu = SP_1+\Omega_{\mathrm{rot}}$. 
The large separation agrees with the dominant spacing for the stars rotating at intermediate rate. 
The large separation for sample stars with the higher mean density and fast rotation agrees 
with $SP_1+\Omega_{\mathrm{rot}}$ and for the stars with lower mean density and slow rotation 
agrees with $SP_2-\Omega_{\mathrm{rot}}$ (if two spacings were found; otherwise the only spacing 
was used in the calculation).

The consistent interpretation of our results using the physical parameters of the targets and 
the agreement with the theoretically expected relation suggest that the unexpectedly large number 
of echelle ridges represents the pulsation and rotation of our target, and not frequencies 
accidentally located along the echelle ridges.
Although we could not reach at this moment the mode identification level using only the 
frequencies obtained from space data, 
this step in disentangling the pulsation-rotation connection is very promising.

The huge database obtained by space missions (MOST, CoRoT and {\it Kepler}) allows us 
to search for regular spacings in an even larger sample and provide more knowledge on how to 
reach the asteroseismological level for $\delta$ Scuti stars.

\acknowledgments{
This work was supported by the grant: ESA PECS No 4000103541/11/NL/KLM. 
The authors are extremely grateful to the referee for encouraging us to include the rotation 
(if possible) in our interpretation. The other remarks are also acknowledged.
}

\end{document}